\documentclass[a4paper, 10pt, titlepage, oneside, onecolumn]{article}
\usepackage[british]{babel}
\usepackage{color}
\usepackage{graphicx}
\usepackage{wrapfig}
\usepackage{lscape}   %Used to for landscape figures
\usepackage{rotating} %Used to for landscape figures
\usepackage{epstopdf}
\usepackage{pdflscape} % Used to ensure PDF pages appear properly in landscapte
\usepackage{amsmath,amsfonts,amssymb}
\usepackage{makeidx}
\usepackage{ifthen}
\usepackage{multicol}
\usepackage{fancyhdr}
\usepackage{enumerate}
\usepackage{subfig} % Supercedes subfigure
\usepackage{wrapfig}
\usepackage[disable]{todonotes}
\usepackage{units}
\usepackage{tabu}
\usepackage{titlesec} %Format section headings
\usepackage{csquotes}
\usepackage{tabularx} % Formatting tables
\usepackage{mdframed} %Used for drawing boxes around text
\usepackage{textcomp} % Generates Euro sign
\usepackage[T1]{fontenc}
\usepackage{lmodern}
\titleformat{\section}{\normalfont\Large\bfseries}{\thesection}{1em}{}
\titlespacing*{\section}{0em}{2\baselineskip}{0\baselineskip}

\titleformat{\subsection}{\normalfont\large\bfseries}{\thesubsection}{1em}{}
\titlespacing*{\subsection}{0em}{1\baselineskip}{0\baselineskip}

\titleformat{\subsubsection}{\normalfont\normalsize\bfseries}{\thesubsubsection}{1em}{}
\titlespacing*{\subsubsection}{0em}{1\baselineskip}{0\baselineskip}

\titleformat{\paragraph}[runin]{\normalfont\normalsize\bfseries}{\theparagraph}{1em}{}

\definecolor{contents_colour}{rgb}{0.4, 0.1, 0.8}

\definecolor{author_colour}{rgb}{0.8, 0.1, 0.4}

\definecolor{tbc_colour}{rgb}{0.4, 0.1, 0.8}

\usepackage[bibstyle=numeric-comp, citestyle=numeric-comp,  minnames = 1, maxnames = 10, giveninits = true, url=true, doi=true, autocite=superscript, sortlocale=english, sorting = none, backend=biber]{biblatex}
\renewcommand{\cite}[1]{\autocite{#1}} %Redefine \cite to be \autocite to allow easy change of citation style

\usepackage[printwatermark]{xwatermark}
\usepackage{xcolor}
\usepackage{graphicx}
\usepackage{lipsum}

\newcounter{NumRec}
\mdfdefinestyle{RecomendationFrame}{%
leftmargin = 20 ,%
rightmargin = 20 ,%
innertopmargin = \topskip ,%
splittopskip = \topskip,
innerleftmargin=10,
innerrightmargin=10,
roundcorner = 30pt,
rightline=true,
frametitlerule=false,frametitlerulecolor=gray,
backgroundcolor = blue!20,%
outerlinecolor = black,%
frametitlebackgroundcolor=lightgray,
frametitlerulewidth=10pt,
nobreak=true}

\newenvironment{recommendation}% environment name
{% begin code
\par\vspace{\baselineskip}\noindent

\begin{mdframed}[style=RecomendationFrame,frametitle=Recommendation \arabic{NumRec}]
\begin{itshape}% 
}%
{% end code
  \vspace{\baselineskip}
  \end{itshape}
  \end{mdframed}
  \ignorespacesafterend
  \stepcounter{NumRec}
} %Input formatting commands for the document

\usepackage{siunitx}

\begin{document}

% Title page
\begin{titlepage}
\begin{center}

{\Huge Plasma Wakefield Accelerator Research 2019--2040\\}
\vspace{5mm}
{\Large \textit{A community-driven UK roadmap compiled by the\\Plasma Wakefield Accelerator Steering Committee (PWASC)}\\}
\vspace{5mm}
{\Large March 2019\\}

\includegraphics[width=180mm]{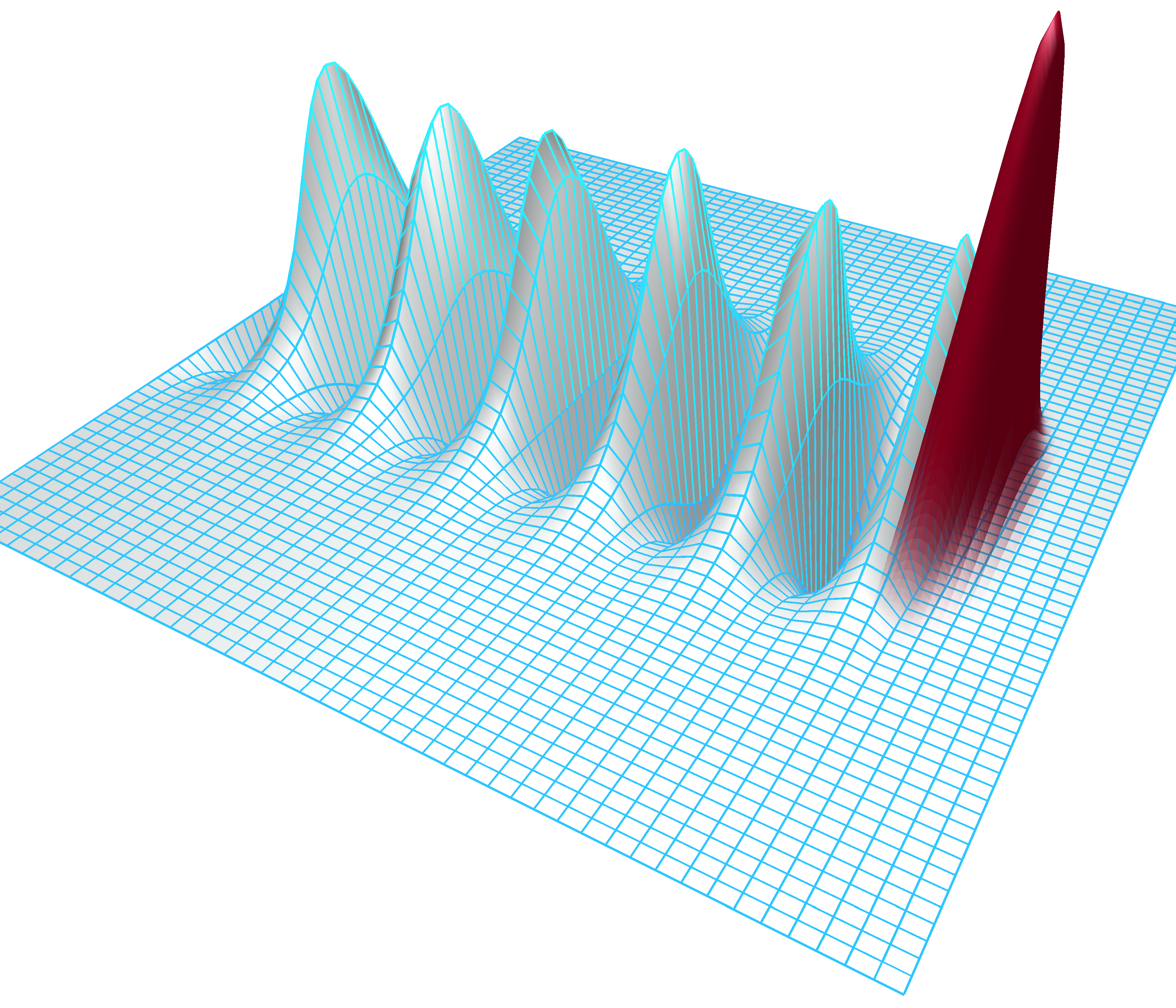}
\end{center}

\noindent{\large Bernhard Hidding, Simon Hooker, Steven Jamison, Bruno Muratori, Christopher Murphy,\\ Zulfikar Najmudin, Rajeev Pattathil, Gianluca Sarri, Matthew Streeter, Carsten Welsch,\\ Matthew Wing, \& Guoxing Xia \\}
\end{titlepage}

% Table of contents page
\pagenumbering{roman}
\tableofcontents
\newpage

% Executive summary
\pagenumbering{arabic}

\section{Executive summary}
The acceleration gradients generated in a laser- or beam-driven plasma wakefield accelerator are typically three orders of magnitude greater than those produced by a conventional accelerator. Plasma accelerators can therefore open a route to a new generation of very compact accelerators in a technological transformation comparable to that enabled by the switch from bulky vacuum tubes to transistors. In addition, plasma-based accelerators can generate beams with unique properties, such as tens of kiloamp peak currents, attosecond bunch duration, ultrahigh brightness and intrinsic particle beam-laser pulse synchronization.

Plasma wakefield accelerators have already generated electron beams with GeV energies --- comparable to the beam energy used in a stadium-sized synchrotron --- from accelerator stages as short as a few centimetres. These beams have been used to produce femtosecond-duration pulses of radiation from visible to hard X-ray wavelengths which themselves have  been used, for example, to make tomographic images of biological samples and to capture shock propagation in laser-shocked materials.

It is clear that plasma accelerators have the potential to be a disruptive technology with broad impact in science, medicine and technology. The ability to drive multiple, synchronised sources of energetic, high-brightness particles and X-ray or $\gamma$-ray photons from compact machines will enable new opportunities in ultrafast science and offers the potential, for example, to provide medical diagnosis and treatment from a single, laboratory-scale facility. Generation of neutral electron--positron plasmas offers the prospect of recreating  extreme astrophysical environments and the study of these spectacular events in the laboratory. In the longer term, plasma accelerators could provide a way to realise high-energy physics colliders required at the forefront of physics.

These transformative prospects motivate the rapidly increasing number of laboratories around the world which now engage in plasma accelerator R\&D, with major projects in Europe, the USA, China, Korea, and Japan. The UK groups have very high international standing in this field, which is based on an impressive, decades-long record of achievement. UK groups continue to be at the forefront of new developments in plasma accelerators, and for this reason UK scientists play major roles in almost all of the international research projects in this field. Nevertheless, this position is threatened by a known lack of investment in national and university-scale research facilities, and by the increase in research investment by other countries.

Plasma accelerator research in the UK grew out of pioneering research by university groups, in partnership with national laboratories. As plasma accelerators mature and move from being the object of study to a driver of applications, as the number of UK groups working in the field has grown, and as merging  with conventional accelerator science and the influence on other fields of science picks up pace, it has become clear that work on plasma accelerators requires more coordination. The research community has responded to this need by the establishing the UK-wide Plasma Wakefield Accelerator Steering Committee (PWASC, \href{http://pwasc.org.uk/}{http://pwasc.org.uk/}) to help coordinate the activities of key stakeholders. Members of PWASC are drawn from UK research groups at universities, the Central Laser Facility, the Accelerator Science and Technology Centre, and the two Accelerator Science Institutes.

The need for a roadmap for plasma accelerator research was identified, a draft roadmap was produced by members of PWASC and other members of the UK community; feedback and further input on this was gathered at a Community Meeting held in January 2018, and feedback on a second full draft was obtained via email at the end of 2018.  Community members also contributed to STFC's 2017 Accelerator Strategic Review Report, the consultation period for which coincided with the development of the present roadmap. While the STFC report covers all aspects of accelerator science, the present roadmap provides a more detailed consideration of plasma accelerator research. It should also be emphasised that this roadmap focuses on \emph{plasma wakefield} acceleration and does not discuss in detail very important, and closely related, fields such as ion acceleration and dielectric accelerators.

In the sections below we review the state of the art, outline potential applications, describe the research and development required to enable those applications, and discuss synergies with related areas of research. We also set-out the resources required to realise these ambitions and provide a timeline for advances in the key areas.

This roadmap contains the following recommendations, with the ordering as they appear in the main text and no ranking implied.

\pagebreak
\subsection*{Recommendations}

\begin{enumerate}
\item  \hyperref[Rec:CLF]{A facility with target areas and beamlines dedicated to research on laser- and hybrid laser-driven plasma accelerators and their applications} should be developed at CLF; these facilities should be state of the art in terms of beam stability, and should take advantage of the UK's lead in high-average power laser technology to allow operation at repetition rates above \unit[10]{Hz}.

\item \hyperref[Rec:UniversitySystems]{The development and operation of UK university-scale labs} for laser-plasma accelerator research and applications, including e.g. SCAPA beamlines, should to be supported and exploited.

\item \hyperref[Rec:EuPRAXIA]{The UK should aim to play a key role, and support the pan-European EuPRAXIA  initiative} for a European Plasma Research Accelerator with eXcellence In Applications.

\item \hyperref[Rec:CLARA]{A dedicated plasma wakefield acceleration beamline} should be developed at CLARA/ASTeC.

\item \hyperref[Rec:AWAKE]{The UK should provide substantial, and increased, investment in the AWAKE Run 2 (2021--4) programme.}

\item \hyperref[Rec:ProgrammaticFunding]{Programmatic funding} of plasma wakefield accelerators should be provided to optimise the quality and stability of plasma-accelerated particle bunches, to increase pulse repetition rates, and to enable key applications in the industrial sector and in fundamental science.

\item \hyperref[Rec:Cross-Council]{UKRI should develop mechanisms for providing cross-council support} for the wide range of research, in a variety of settings,  necessary to drive advances in plasma accelerators; this range includes fundamental research (e.g.\ plasma physics), technology development (e.g.\ novel lasers), and application development (e.g.\ medical imaging).

\item \hyperref[Rec:IntlCollaborators]{Mechanisms should be sought to allow UK groups to play leadership roles in international high-visibility collaborations} such as SLAC FACET-II, Helmholtz ATHENA, Laserlab Europe, ELI, ARIES etc., and to exploit these.

\item \hyperref[Rec:NationalScheme]{A national scheme should be developed to enable mobility and knowledge transfer within UK institutions}, to increase beam access, and to sustain collaborative efforts; this would provide a means to test new concepts, train students, and prepare for beam time at national and international facilities.

\item \hyperref[Rec:Fellowships]{A new "Novel Accelerator Fellowship" scheme} should be developed and the support available for training PhD students in novel accelerators should be increased.
\end{enumerate}

\newpage
\section{Motivation and introduction}
Particle accelerators play a central role in numerous areas of science and technology \cite{web:EUCARD2ApplicationsAcceleratorsEurope}. They have been instrumental in key discoveries in particle physics that have enabled our understanding of the building blocks of the universe, and in this sphere alone they have played a vital role in discoveries which led to tens of Nobel prizes at a rate of approximately one every 3 years \cite{Haussecker:fh}. Accelerators drive the latest generation of light sources, which deliver unprecedented information about the structure of materials from  the structure of complex proteins, to tumour identification, and understanding the function of high performance materials. In medicine, particle accelerators are used for imaging and treatment, as well as to generate radio-isotopes. In the realm of national security, particle accelerators are used for cargo inspection and in airport scanners.

\begin{figure}[tb]
    \centering
    \includegraphics[width = 100mm]{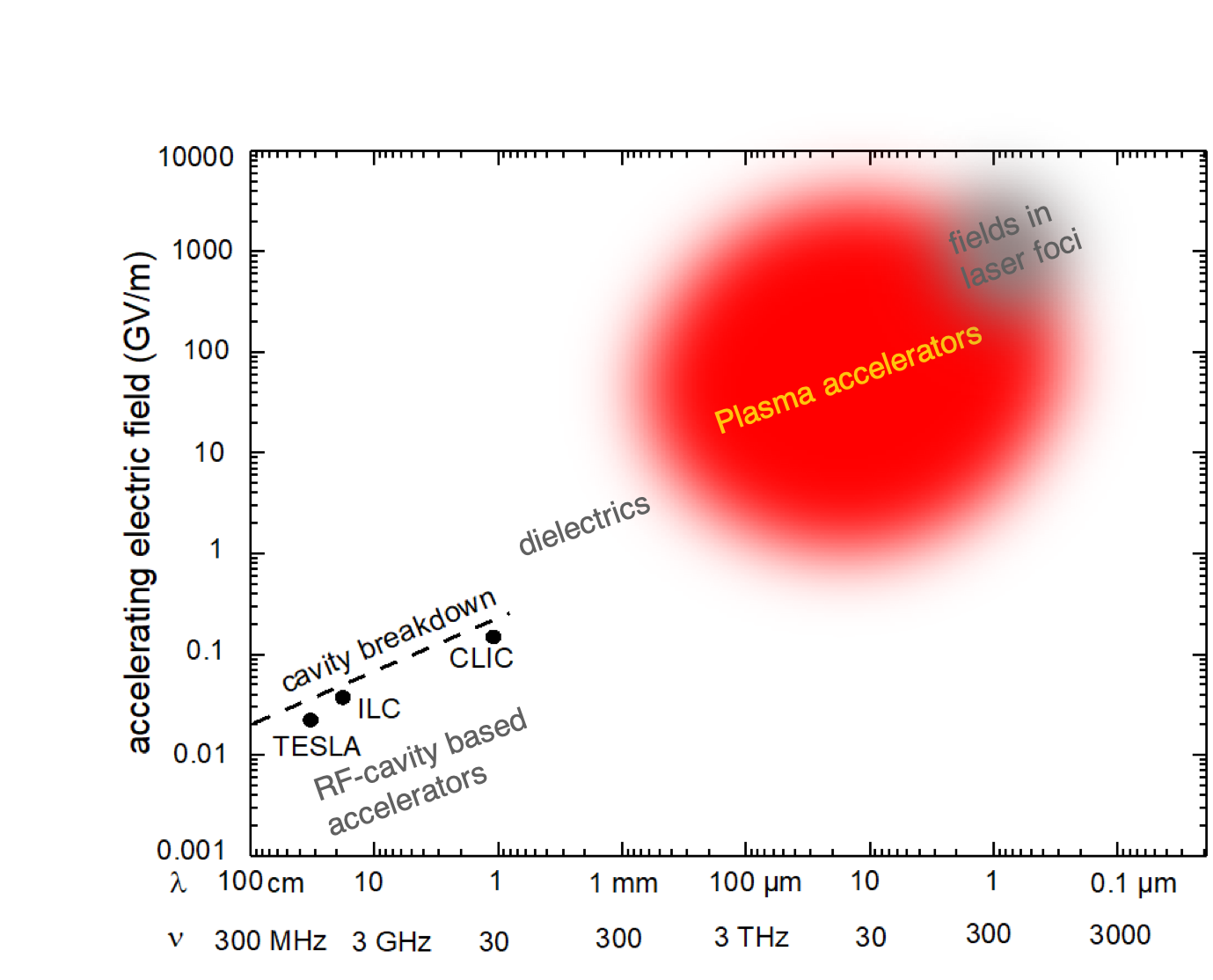}
    \caption{Field gradients of different accelerator technologies and their relation to wavelength and frequency of the accelerating field.}
    \label{Fig:Field_gradient}
\end{figure}

Plasma-based accelerators are important because they can accelerate particles to high energies in distances a thousand times shorter than conventional radio-frequency (RF) accelerators (see Fig.\ \ref{Fig:Field_gradient}). This fact alone promises a technological transformation comparable to that enabled by the switch from bulky vacuum tubes to transistors. Plasma-based accelerators are also important because they generate beams with unique properties, such as high peak currents, femtosecond to attosecond bunch duration, low emittance and ultrahigh brightness. They therefore offer the prospect of a new generation of very compact accelerators with several potential near-term applications, such as driving next generation light sources or studying high field effects; in the longer term they may offer a route to the beam energies required for future particle colliders.

This roadmap has been prepared by the UK Plasma Wakefield Accelerator Steering Committee (PWASC), with input from all those working in the UK on this topic (Appendix \ref{Sec:Community_consultation} describes how input from the community was obtained). In this roadmap we review the state of the art,  outline potential applications,  describe the research and development required to enable those applications, and discuss synergies with related areas of research. We also set-out the resources which would be required to realise these ambitions and provide a timeline for advances in the key areas. The roadmap concludes with recommendations for research funding, provision of experimental and computational facilities, training, and international collaboration.

\textbf{It is important to emphasise that the remit of PWASC is \emph{plasma wakefield} acceleration. As such the roadmap focuses on this area and does not discuss in detail very important, and closely related, synergistic fields such as ion acceleration and dielectric accelerators.}

\section{State-of-the-art}
\subsection{Overview}
Plasma wakefield acceleration can be driven by intense laser pulses or by particle bunches. In contrast to conventional linacs, where a series of static cavities accelerate particles, plasma-based cavities co-propagate with the driving pulses with approximately the speed of light. There are several flavours of plasma wakefield acceleration, each with its own advantages and challenges. This variety allows many of the major goals of accelerator R\&D, and their applications, to be addressed. The transformative prospects of plasma acceleration are reflected by the increasing number of labs around the world, including national facilities, which now work on plasma acceleration, as summarised in Fig.\ \ref{fig:WorldwidePlasmaLabs}.

There are two major approaches of plasma wakefield acceleration: laser-driven plasma wakefield aocceleration and particle-beam driven plasma wakefield acceleration. These are discussed briefly below.

\subsection{Laser wakefield accelerators (LWFAs)}
In laser wakefield accelerators (LWFAs) the plasma wave is driven by one or more intense pulses of laser radiation. Laser wakefield acceleration has made rapid technical and conceptual progress, not least since the milestone demonstrations of quasi-monoenergetic beams, the first generation of GeV-scale beams and first LWFA-based undulator radiation  --- all examples of world-leading breakthroughs made with significant involvement from UK groups \cite{Mangles:2004, Leemans:2006,Schlenvoigt:2008}. The driving laser systems are reasonably compact, and (for low pulse repetition rates) are commercially available. It is also relatively easy to inject electrons from the plasma itself, and to accelerate them to energies of tens of MeV to a few GeV in gas jets, gas cells, or plasma channels.

A defining feature of LWFAs is that the driving laser pulse propagates at its group velocity, which is slightly below $c$. As such relativistic particle bunches gradually overtake the plasma wakefield and eventually move into a decelerating phase, a process known as ``dephasing''.

In most regimes the dephasing length is much longer than the distance over which the laser pulse naturally diffracts, i.e.\ the Rayleigh range. For example, for laser focal spot sizes in the range $\unit[10 - 100]{\mu m}$, the Rayleigh range is \unit[0.3 -- 30]{mm}. In comparison, a \unit[10]{GeV} accelerator requires a plasma length of around \unit[1]{m}.

Broadly speaking LWFAs operate in one of two regimes. In the quasi-linear regime the plasma wave is approximately sinusoidal which means that it can provide nearly equal size phases for accelerating and focusing particles of positive or negative charge. In this regime the driving laser pulse must be guided over the length of the accelerator by an external waveguide. To date, guiding by grazing-incidence reflections in a hollow capillary or gradient refractive index guiding in plasma channels has been employed. LWFAs driven in \unit[200]{mm} long plasma channels have been used to reach an  electron energy of \unit[7.8]{GeV}. \cite{Leemans:2014lka, Gonsalves:2019ht}

In the highly non-linear, or ``bubble'', regime the laser intensity is so high that essentially all the plasma electrons are expelled from the region immediately behind the driving laser pulse. In this regime the laser pulse is self-guided by the relativistic response of the plasma and the transverse density gradients within the bubble structure. Self-guiding over tens of millimetres of plasma to generate few-GeV electron beams has been demonstrated.

The particle bunches generated by LWFAs have pulse durations of order a few femtoseconds, charges in the range \unit[10 -- 1000]{pC}, a normalised emittance of order \unit[1]{mm mrad}, and a relative energy spread in the range \unit[1 -- 10]{\%}. However, since in most work to date the electrons are injected into the wakefield by stochastic processes,  the shot-to-shot jitter in these parameters can be high. There is therefore considerable effort underway to develop methods for controlling the processes by which electrons are injected and trapped in the plasma wave; the ability to control the injection process will not only reduce shot-to-shot jitter, but also improve (and enable control of) the bunch parameters.

Owing to dephasing, diffraction and/or energy depletion of the laser driver, it is likely that multiple LWFAs will need to be coupled together to reach particle energies much in excess of \unit[10]{GeV}. In order to reduce the length of the optical system used to couple in the driving laser pulse into each stage, tape-based plasma mirror systems have been developed. In these, a thin tape is placed between the two stages; particles from the previous stage pass through this, but the drive pulse for the second stage is reflected into the second stage by the plasma it forms on the surface of the tape. It has been demonstrated that this approach --- together with the use of compact active plasma lenses, based on capillary discharges  --- can couple two  \unit[100]{MeV}-scale plasma stages in a distance of a few centimetres (see \S \ref{Sec:Transport_staging_feedback}).

\subsection{Plasma wakefield accelerators (PWFAs)}
In plasma wakefield accelerators (PWFAs) the driver is an intense bunch of particles. An important feature of PWFAs is that acceleration can be maintained over a long distance since, unlike with a laser driver, dephasing is not relevant (for relativistic electron beam drivers) and the driving beam does not diverge strongly. Further, building on decades of development of \emph{conventional} accelerators, high-energy particle beam drivers are available which can  operate at high repetition rates, with excellent stability, and with good wall-plug efficiency. Indeed, to their mutual benefit, advances in conventional accelerators provide a technological push which is matched by the scientific pull provided by the demands of PWFAs.

A distinct challenge faced by PWFAs is that the rate of progress is limited by the small number of facilities available since large scale conventional accelerators and complex infrastructure required. 

PWFAs can be driven by bunches of electrons, positrons, or protons. Experiments with electron or positron drivers have been undertaken at several facilities, mainly in the USA. Electron-driven PWFAs are the most prevalent type of PWFA since its inception \cite{Chen1985PhysRevLett.54.693} and first demonstration  \cite{Rosenzweig:1988}. For example, in 2007 experiments at SLAC demonstrated doubling the energy of a \unit[42]{GeV} electron beam \cite{Blumenfeld2007}, and efficient acceleration of a separate witness bunch has been demonstrated at the same facility \cite{Litos2014Natureshort}. Positron beams have been used not only as drivers, but also have been accelerated  \cite{cordepositronsnature2015,GessnerNatComm2016}. Complemented by further programmatic R\&D on fundamental beam-plasma interaction studies, this fuels prospects for a potential path towards a high energy physics collider, next to various key applications such as advanced light sources. 

More recently a hybrid approach has been proposed which aims to combine many of the advantages of LWFAs and PWFAs. In the hybrid LWFA $\rightarrow$ PWFA, a LWFA generates an electron bunch which then drives a subsequent PWFA stage. This provides a very compact particle beam driver, with inherent synchronization between the laser and electron pulses. Additional laser beams can be used for preionization, injection, manipulation and plasma and bunch diagnostics.

Proton beams as drivers for PWFAs were considered as early as 1986 \cite{Katsouleas1986}. In 2009, a more thorough theoretical study investigated the use of LHC-scale proton beam drivers to generate TeV-scale beams \cite{Caldwell:2009}. The CERN Advanced Wakefield Experiment (AWAKE) \cite{web:awake} has made huge steps forward, and very recently this approach demonstrated acceleration of an electron witness bunch for the first time \cite{Adli:2018jz}.  At present the proton bunches available are much longer than the plasma period, and hence it is necessary to modulate them by seeding a self-modulation process. Ultimately it may be possible to avoid this step by generating short proton bunch drivers. Other challenges for p-PWFAs are the development of long ($> \unit[10]{m}$) plasma cells, and the development of short-bunch  electron injectors.

Current research into PWFAs is focused on: controlling injection and trapping of particles into the wakefield, e.g. by plasma cathodes, photocathodes and related approaches \cite{hiddingTrojanpatent2011, Hidding:2012, WittigPRSTAB2015}; improving the output beam quality and stability; and reducing the footprint of the drivers.

\section{National and international perspectives}
\subsection{The international perspective}
\begin{figure}[tb]
\centering
\includegraphics[width=140mm]{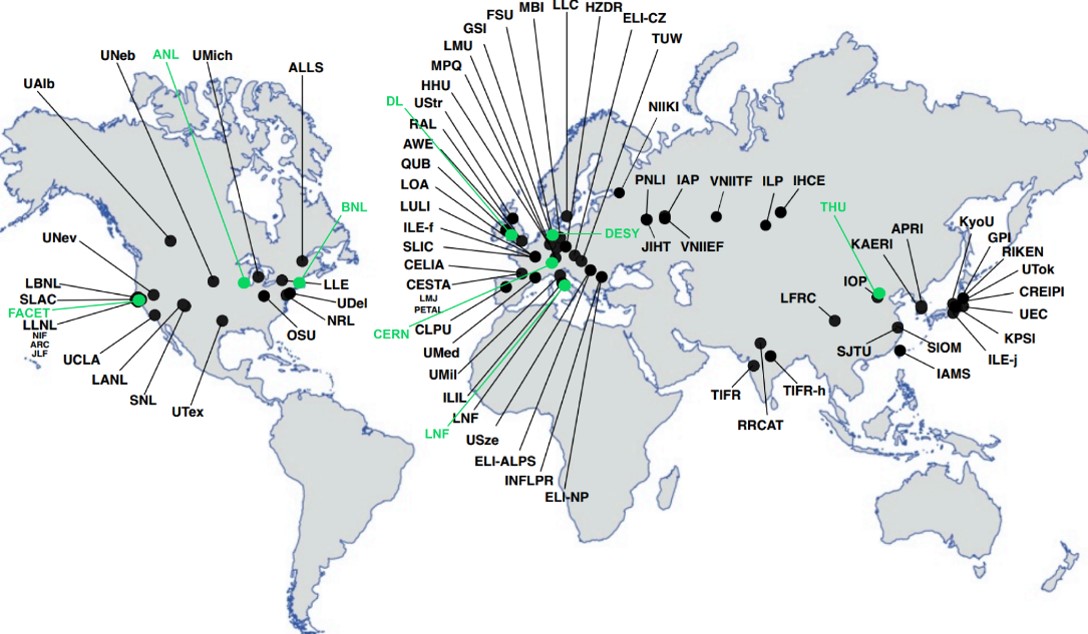}
\caption{Non-exhaustive overview of laboratories working on (or with the capacity to work on) laser-driven (black) and particle beam-driven (green) plasma wakefield R\&D. Based in part on the map of high-power laser laboratories produced by the International Committee on Ultra-high Intensity Lasers (ICUIL).\cite{web:ICUIL_lasers}}
\label{fig:WorldwidePlasmaLabs}
\end{figure}

Figure \ref{fig:WorldwidePlasmaLabs} shows laboratories around the world working on experimental plasma wakefield acceleration. The growth of this research area is evidenced by the fact that many of these laboratories started work in this field only recently.

\subsubsection{LWFAs}
Today, LWFAs constitute the largest part of the  research worldwide on plasma accelerators, partly due to the availability of compact high-power laser systems (typically Ti:sapphire lasers) and the relatively low technological threshold for initiating research programmes in this area.  There are many multi-institutional national and international projects and initiatives devoted to LWFAs.  

LWFA research is very strong in Europe. Here, Laserlab Europe \cite{web:laserlab} plays an important role in enabling campaign-based access to high-power laser facilities across Europe; EuPRAXIA \cite{web:eupraxia} is an EU H2020 project across LWFAs and e-PWFAs to design a European facility based on plasma accelerators; and the Extreme Light Infrastruture (ELI) \cite{web:eli} is a European Strategy Forum on Research Infrastructures (ESFRI) project for the investigation of light-matter interactions at the highest intensities and shortest time scales.  There are particularly strong research groups based in Germany, France, Italy, and Portugal.

The USA has a large number of groups working on LWFAs. Of these, the biggest is the BELLA programme based at Lawrence Berkeley National Laboratory (LBNL), but there is a large number of other significant research groups based at other national laboratories and in universities.

There is significant and important research on LWFAs undertaken by the large number of groups based at national facilities and university laboratories in China, India, Korea and Japan.

\subsubsection{PWFAs}
Research on PWFAs was initially restricted to a handful of large research centres, but the intensity of research is now strongly ramping up as more laboratories engage. In the field of e-PWFAs, there are several important multi-national collaborations providing programmatic or campaign-based R\&D at, for example, SLAC FACET(-II), BNL ATF(-II), DESY FLASHForward, INFN SPARC-X and elsewhere.   Research on p-PWFAs, through the AWAKE experiment, is strongly supported by CERN and involves a  large collaboration by the standards of plasma wakefield acceleration. As mentioned above, the EuPRAXIA project includes a programme on linac-driven plasma wakefield accelerator concepts.

\subsection{The UK perspective}
\subsubsection{LWFAs}
The UK has several internationally leading groups; these are mostly university-based, several of which are also affiliated with one of the two Accelerator Institutes.  The UK groups have made major contributions to fundamental research on LWFAs. These include the first demonstration of the generation of narrow energy spread beams;\cite{Mangles:2004} pioneering demonstrations of acceleration to the GeV range in externally-guided\cite{Leemans:2006} and self-guided\cite{Kneip:2009} geometries; the development of novel plasma channels\cite{Spence:2000fr, Butler:2002, Wiggins:2011kr}; studies of novel methods for controlling electron injection via ionization of dopant species\cite{Rowlands-Rees:2008, Bourgeois:2013ez}; and  measurements of the duration \cite{Heigoldt:2015} and emittance\cite{Brunetti:2010} of laser-accelerated electron bunches. UK groups have also played leading roles in demonstrating applications of LWFAs, including: the generation of visible and extreme ultraviolet undulator radiation from laser-accelerated electrons;\cite{Schlenvoigt:2008, Fuchs:2009} the generation of bright betatron radiation with photon energies in the \unit{keV} to \unit{MeV} range\cite{Kneip:2010, Cipiccia:2011}; the application of betatron radiation to tomographic imaging of human bone \cite{Cole:2015fh}; and applications to fundamental physics, such as studies of the radiation reaction \cite{Cole:2018kp}.

To date most experimental work by the UK groups has been performed at the Central Laser Facility (CLF) at the Rutherford Appleton Laboratory (RAL), or at laser facilities outside the UK.   The Astra-Gemini TA3 laser, commissioned in 2008 at RAL, was a major advance for laser driven particle acceleration in the UK. This is not only because it features an ultrashort high-intensity pulse, which is ideal for laser wakefield acceleration, but also because it operates at relatively high-repetition rate of 1 shot every 20 seconds, compared to the few shots per hour of previous petawatt-scale laser facilities. The Gemini laser increased by a factor of 10 the laser energy provided by the Astra TA2 laser which was used in the first demonstration of the generation of monoenergetic, self-injected beam experiments \cite{Mangles:2004}. The increased pulse energy available from the Gemini laser allows a laser wakefield to be driven to close to wavebreaking at lower density, increasing the phase velocity of the wakefield, and thereby allowing electrons to reach a higher energy before being dephased.

The CLF is heavily oversubscribed, partly due to its reliance on a single, multi-purpose target area, which limits programmatic and cutting-edge R$\&$D in LWFAs. In order to keep international leadership,  a new high-energy, high-repetition rate, state-of-the-art facility is required.

\begin{recommendation}
\label{Rec:CLF}
A facility with target areas and beamlines dedicated to research on laser- and hybrid laser-driven plasma accelerators and their applications should be developed at CLF; these facilities should be state of the art in terms of beam stability, and should take advantage of the UK's lead in high-average power laser technology to allow operation at repetition rates above \unit[10]{Hz}.
\end{recommendation}

Small- and medium-sized laser systems --- such as those based at Imperial College, Oxford, Strathclyde and Queen's University Belfast (QUB) --- play an increasingly important role in allowing groups to develop new concepts, to prepare for experiments at national facilities, and to train students. It is vital that these laser systems are updated and supported in order to have a sustained activity in this area in the UK.

\begin{recommendation}
\label{Rec:UniversitySystems}
The development and operation of UK university-scale labs for laser-plasma accelerator research and applications, including e.g. SCAPA beamlines, should to be supported and exploited.
\end{recommendation}

 The UK groups make substantial contributions to European programmes, such as EuPRAXIA, as well as to other collaborations within Europe, often via joint experiments at Laserlab Europe. They also work with many institutions outside Europe; although these are often funded on an ad hoc basis, these collaborations play an important role in enabling the UK groups to undertake world-class research.
 
 \begin{recommendation}
\label{Rec:EuPRAXIA}
The UK should aim to play a key role, and support the pan-European EuPRAXIA  initiative for a European Plasma Research Accelerator with eXcellence In Applications.
\end{recommendation}

 \subsubsection{PWFAs}
To date the UK groups have contributed to PWFA R\&D mainly via experiments at SLAC FACET and at CERN AWAKE. The CERN AWAKE programme has received funding from the institutes and the STFC since 2012; UK groups play a significant role in AWAKE and constitute approximately 20\% of the authors. At SLAC FACET, the E210 and E203 experiments have been led by UK researchers, UK groups have significant involvement with upcoming PWFA research at DESY,  and in a Europe-wide hybrid LWFA $\rightarrow$ PWFA collaboration. UK groups are the largest fraction of non-US  groups in the currently commencing exploitation period of SLAC FACET-II. This includes leadership on several flagship experiments. A UK-US support pathway is required to exploit these opportunities. The UK groups play a major role in the EuPRAXIA project on LWFAs and PWFAs, representing 6 of the 16 partners and receiving 21\% of the  total funding. 

This multi-year experience, partially obtained without any funding from the UK RC's, demonstrates:

\begin{enumerate}[(a)]
\item There is strong intellectual leadership of UK researchers and institutes on both electron-driven as well as proton-driven PWFAs.
\item There is an urgent need to support these activities with significant funding commitment.
\item There is need to develop experimental PWFA capabilities in the UK, in addition to commitment to international collaboration.
\end{enumerate}

Regarding c), there are promising opportunities in the UK for both electron-linac driven PWFAs as well as for hybrid LWFA $\rightarrow$ PWFA in addition to the partially established R\&D capabilities to support proton-driven PWFAs at CERN in the AWAKE project. 

CLARA, the Compact Linear Accelerator for Research and Applications, offers R\&D opportunities for electron-linac driven PWFAs. CLARA is a dedicated R\&D facility for developing FEL R\&D and to prepare and support UK X-FEL capabilities. CLARA currently produces \unit[50]{MeV} electron beams, and will eventually provide \unit[250]{MeV} beams. While not currently funded, there is the potential to extract the \unit[250]{MeV} beam into a beamline directed to science experiments requiring short pulses of relativistic beams; such a capability may be of interest to the plasma acceleration community.

\begin{recommendation}
\label{Rec:CLARA}
A dedicated plasma wakefield acceleration beamline, with synergies to other novel accelerator schemes, should be developed at CLARA/ASTeC.
\end{recommendation}

As in AWAKE, which relies on self-modulation of the long proton beam, it has been proposed that long electron bunches, such as those use at Diamond, could be modulated.  Using LWFAs to seed the modulation of the Diamond beam leads to short bunches that are radially polarised on creation.  The oscillation of the micro-bunches leads to the generation of radially polarised X-ray pulses of much higher brightness and over a wider photon energy range than the Diamond facility.  This has been studied in detail in simulations and provides an opportunity for an experimental programme which could then lead to upgrades for light sources world-wide.   

The UK is in principle in a strong strategic position with respect to hybrid LWFA $\rightarrow$ PWFA. Such research could be undertaken, for example, at the CLF and at SCAPA. A dedicated  hybrid LWFA $\rightarrow$ PWFA beamline could potentially be implemented at one of the (up to 7) beamlines of SCAPA, thereby allowing programmatic development.  Such programmatic R\&D at laser powers accessible at SCAPA could be complemented by campaign-based R\&D at the CLF with the Gemini laser. Here, the higher laser power levels, and in particular the two-laser beam capability, which is a strong asset for hybrid plasma acceleration, could be exploited, and may then also lead to a dedicated beamline in future CLF upgrades.

In 2018, AWAKE completed data-taking for its Run 1 and achieved its major milestones.  
The experiment demonstrated the self-modulation of the long proton bunches and the 
acceleration of electrons up to \unit[2]{GeV} in the wakefields driven by the 
microbunches.
A future AWAKE programme is being developed in which electrons are expected to be accelerated to about \unit[10]{GeV} over about \unit[10]{m}, with high bunch quality.  This should be a reproducible and scalable process such that first applications of the AWAKE scheme could be considered.  There are significant opportunities for UK contributions to the electron source, plasma technology and general aspects of the experiment such as diagnostics.  The UK should also continue its leading efforts on applications of the AWAKE scheme to high energy physics experiments.          

\begin{recommendation}
\label{Rec:AWAKE}
The UK should provide substantial, and increased, investment in the AWAKE Run 2 (2021--4) programme.
\end{recommendation}

% Scientific and technical Challenges
\section{Scientific and technological R \& D}

\subsection{Plasma drivers and sources}
\subsubsection{Next generation laser drivers}\label{Sec:Next_gen_laser_drivers}
Advances in LWFAs have often been driven by advances in laser technology, and this will continue to be the case.  The application of plasma accelerators to driving useful radiation and particle sources is constrained by the low repetition rate ($< \unit[5]{Hz}$ for a petawatt-class laser) and wall-plug efficiency ($< 0.1\%$) of the Ti:Sapphire driving lasers which are typically used today. These limitations are imposed by the relatively low quantum efficiency of the Ti:sapphire gain medium, and the use of flashlamp pumping, both of which lead to significant deposition of heat that must be removed between laser shots.

Increasing the repetition rate to 10s, 100s, or 1000s of pulses per second requires the development of new laser technologies, and possibly the development of new strategies for driving the wakefield. The UK has made significant contributions in both areas: CLF's DiPOLE project has led to the development of new pump lasers able to provide $\unit[> 100]{J}$ pulses at \unit[10]{Hz}; and UK groups have proposed\cite{Hooker:2014ij} that plasma wakefields could be driven by trains of low-energy laser pulses, potentially enabling the wakefield to be driven by emerging laser technologies.

In order to realise the scientific goals of laser-driven wakefield accelerators mentioned in this roadmap, high power lasers based on advanced technology would be necessary. In particular, to drive applications of laser-based secondary sources, high-repetition-rate, high-energy (including petawatt-class) lasers will be required. 

\paragraph{Diode-pumped petawatt lasers}

 The overall efficiency of laser pump diodes is close to $50\%$, and hence replacing flashlamps with laser diodes would substantially increase the laser efficiency and allow significant increases in repetition rate. This diode pumped solid state laser (DPSSL) technology will therefore bring a paradigm shift in the performance of high power lasers, combining high peak power ($> \unit[1]{kW}$) with high repetition rate (\unit[10]{Hz} -- \unit[1]{kHz}) and high overall efficiency ($>20\%$). Such lasers are being developed at a few places around the world, including STFC's CLF.

A significant part of the work described in this roadmap would rely on a petawatt laser (\unit[30]{J}, \unit[30]{fs}) driven at \unit[10]{Hz} in the short-to-mid term. For future-proofing, it is imperative that this laser is based on DPSSL technology. For example, the DiPOLE100 system developed at CLF would have enough pump power to drive a petawatt laser at \unit[10]{Hz}.

\paragraph{High-peak-power, kilowatt mean power lasers}
Thin-disk and fibre lasers pumped by DPSSLs are being investigated as technologies for driving high-peak-power lasers at \unit[100]{Hz} repetition rates and beyond. For example, CLF is  investigating \unit[100]{TW}, \unit[100]{Hz} lasers; and the iCAN project plans to coherently combine thousands of low-power fibre lasers to generate high-power laser pulses at multi-kHz repetition rates at wall-plug efficiencies significantly above $10\%$.

An alternative scheme is the multi-pulse LWFA (MP-LWFA) approach being developed by UK groups:\cite{Hooker:2014ij} here the driving laser energy is delivered over tens of plasma periods  in a train of low-energy pulses or a long, modulated pulse. Delivering the drive energy over a longer interval (\unit[1 -- 10]{ps}) allows the use of different laser technologies (e.g.\ fibre or thin-disk lasers),  capable of multi-kilohertz repetition rates and high wall-plug efficiencies. Many of these ideas can be tested using laser systems available in university laboratories, and at CLF, and there is a clear  opportunity for the UK to build on its world lead in this area.

\subsubsection*{Mid and far IR lasers}
Today most LWFAs are driven by  lasers operating at wavelengths at around 1 $\mu$m. However, high-power laser systems operating at longer wavelengths ($\lambda = \unit[5 - 10]{\mu m}$) could expand the parameter range of LWFAs considerably \cite{PogorelskyPhysRevAccelBeams.19.091001}.

For example, since the force responsible for driving the wakefield, the ponderomotive force, scales as $I \lambda^2$, increasing the wavelength of the driver allows the laser intensity to be reduced. This in turn allows cold electrons to be injected into the wakefield via tunneling ionization of a dopant species by a low-intensity injection pulse, which could lead to low emittance electron bunches similar to the PWFA-based plasma photocathode or Trojan Horse approach. Long wavelength laser pulses also may be useful to preionise comparably broad plasma channels for PWFA systems.

R\&D on these approaches both conceptually and as regards mid to far IR laser systems such as CO$_2$ lasers is required to assess and develop these scenarios.

\subsubsection{Next generation particle beam drivers}
\label{Sec:Next_gen_beam_drivers}
Intense particle beams are required to excite intense plasma oscillations and to drive intense plasma waves over extended distances.  Short beams with high peak currents are ideally suited to drive resonant plasma wakes into the nonlinear blowout regime; and high energy, low emittance beams  are necessary to realise extended acceleration distances. It is a fortunate synergy that substantial R\&D has  been put into the development of photocathodes and advanced compression schemes  in order to drive hard X-ray free-electron lasers such as the LCLS and European XFEL. Nevertheless, further optimization is necessary to produce an ideal  driver for PWFAs: for example, a high peak current is of even higher importance for PWFAs than for FELs; whereas, in contrast, a small energy spread is crucial for a FEL, but is less important for driving a PWFA stage.   

Trapping of electrons from the background plasma requires high peak driver currents $> \unit[6]{kA}$, although this can be reduced to  $\unit[2]{kA}$ or less using a density downramp, plasma torch downramps \cite{WittigPRSTAB2015}, or a downramp-assisted plasma photocathode. Dedicated efforts are now underway at SLAC FACET-II to produce beams from linacs with peak currents exceeding tens of kA. \cite{web:facetIITDR}

The optimum parameters of the driver depend strongly on the density of the plasma. Lower density plasmas can be driven by longer drive beams, likely improving the shot-to-shot stability and uncorrelated energy spread of the bunch. On the other hand, very short (few or sub-fs) electron beam drivers could drive $\unit{TV\,m^{-1}}$ stages at high plasma density. 

The ability to shape electron beams from photocathode linacs is also important. For example, triangular bunch current shapes can be used to harness higher transformer ratios, or to develop reduced energy spreads via direct beam loading. Other parameters of the driver which require optimization include energy spread and emittance; finding the best trade-off between these, and other, parameters of the driver requires dedicated machine development time. We note that CLARA offers outstanding opportunities for develping electron PWFA drivers since it offers a state-of-the-art photocathode with the potential to operate at a repetition rate up to \unit[400]{Hz}.

Finally, high repetition rates are important, which can for example be produced by superconducting linac technology. More research will then also be required to explore the plasma response at such high repetition rates.

\paragraph{Proton-driven PWFAs}
A special approach is required for proton-driven PWFAs since short, intense high energy proton beams are not readily available from linacs. Here the approach is to exploit self-modulation of long proton bunches, such as those produced at the LHC. 
The possibility of pre-modulating the long proton bunches or generating shorter drive bunches are areas of interest for proton-driven PWFAs.
plasma.

\paragraph{Hybrid LWFA--PWFA}
Another possible next-generation driver could be electron bunches generated  from a LWFA. These are particularly attractive for PWFAs since they are of short duration and have peak currents exceeding tens of kA \cite{Couperus2017}. Although LWFA bunches often have high relative energy spread (tens of percent), for driver bunches $\gtrsim \unit[100]{MeV}$ all the electrons propagate at a speed close to $c$. This hybrid LWFA--PWFA allows PWFAs to be achieved with any LWFA system and offers several advantages, including:  elimination of dephasing; and excellent laser--electron bunch synchronization.

Significant progress has been made with this approach\cite{web:radiabeamAAR, web:eupraxia}, much under UK leadership. However, it has received no significant UK research council funding. Substantial R\&D is required to explore this area, which could be undertaken at UK laser centres  such as SCAPA, CLF, and university-based labs.

\subsubsection{Plasma sources}
A core component of a plasma accelerator is the plasma itself. For both PWFAs and LWFAs this must comprise a region of plasma in which the species, ionization state, density, uniformity, and length are all well defined.  Since plasma does not exist at room temperature, the plasma must be created by the driving beam itself, or by auxilliary beams or electrical discharges.

For more advanced plasma accelerators it may be necessary to control the longitudinal and/or transverse profile of the plasma density. For LWFAs, control of the longitudinal profile allows the possibility of overcoming dephasing, increasing the energy gain per stage. The development of hollow, or near-hollow, plasma channels is of significant current interest for LWFAs and PWFAs since the focusing plasma wakefields are weak, which prevents emittance growth. Hollow channels could be particularly important for positron acceleration since the accelerating fields have a larger amplitude, and occupy a greater proportion of the plasma wave, than those generated in uniform plasma.

Some requirements of the plasma source are specific to the driver. For laser-driven plasma accelerators the intensity of the driving laser is nearly always more than sufficient to ionise a target gas to create the plasma and hence auxiliary ionization systems are not usually required. For LWFAs the length of the plasma source should match the shorter of the dephasing or pump-depletion lengths. Alternatively, in advanced schemes, dephasing can be overcome partially by controlling the longitudinal profile of the plasma density. The length of the transition region between the body of the plasma and the surrounding vacuum is important for laser-driven plasma accelerators since it determines the extent to which the laser is defocused as it is coupled into the plasma.  Finally, in the quasi-linear regime it is necessary to guide the driving laser pulse over the length of the accelerator stage using an external waveguide capable of withstanding laser intensities of order $\unit[10^{18}]{W\,cm^{-2}}$. 

Particle-driven plasma accelerators typically require longer (up to several metres), lower-density ($n_\mathrm{e} \approx \unit[10^{14} - 10^{18}]{cm^{-3}}$) plasma than the laser-driven case. For PWFAs it may  be necessary to use an auxiliary laser to ionise the source species since the electric fields of the driver may be too low for field ionization. On the other hand, the design of the plasma source in this case is simplified by the fact that neither defocusing of the driver or dephasing occurs to a significant extent.

For both LWFAs and PWFAs, important practical considerations include the operating lifetime and shot-to-shot reproducibility of the plasma source. In some cases, it will also be important to be able to provide diagnostic access to the plasma, or to isolate it from other parts of the beam line and/or vacuum system.

The development of practical plasma accelerators will require a parallel development of plasma sources. For LWFAs self- or external-guiding of higher pulse energies, through longer lengths of lower density plasma is needed to increase the energy gain per stage above \unit[10]{GeV}. The development of hollow plasma channels could be crucial for acceleration of positrons. For PWFAs driven by modulated proton beams, the development of long, uniform plasma cells will be vital. 

As the energy gain of the accelerator is increased, the plasma target will need to handle large particle beam or laser pulse energies without damage. For plasma accelerators to drive applications, the plasma source will need to be reproducible and highly reliable; and as the repetition rate is increased this will need to be the case for millions of shots per day.

Plasma sources play a central role in both laser- and particle-beam-driven plasma accelerators, and their continued development is necessary and important if plasma accelerators are to drive applications. The development of plasma sources should therefore be recognised by the research councils as a priority research area.

\subsubsection{Positron sources}
Future plasma-based electron--positron colliders will require a source of high-quality positron bunches for injection into the plasma wakefield. Although great strides have been made in plasma-based acceleration of electrons, acceleration of positrons in plasma accelerators is still in its infancy. One reason for this is that only one RF facility in the world (FACET-II) can provide positron beams of sufficient quality.

A potential solution to this problem is the generation of high-quality, ultra-relativistic positrons  by interaction of a laser-accelerated electron beam with a solid target. In a nutshell, the positrons are generated as a result of a quantum cascade initiated by a laser-driven electron beam propagating through a high-Z solid target. For sufficiently high electron energy and thin converter targets, the source size, divergence and duration of the generated positron bunches resemble those of the parent electron beam. The positron beams generated this way are therefore attractive for injection into  further plasma accelerator stages, with the important bonus that they are naturally synchronised with a high-power laser system. UK-led groups have recently used this approach to  demonstrate the generation of fs-scale, narrow divergence positron beams \cite{Sarri:2013,Sarri:2015}, together with the first ever generation of a neutral matter--antimatter plasma in the laboratory \cite{Sarri:2015,Warwick:2017}.

UK researchers initiated this area of research and retain a world-leading position, and as a consequence they currently play a significant role in developing positron sources for the EuPRAXIA and ALEGRO projects. 

Further optimization of these sources will require the charge  of the positron bunch to be increased and its  emittance to be decreased, both of which can be achieved by increasing  the charge and energy of the parent electron beam. 

We note that the generation of neutral electron--positron plasmas presents an opportunity to recreate in the laboratory extreme astrophysical environments such as the atmosphere of quasars and active galactic nuclei. The physical understanding of these environments is based on astronomical observation and numerical modelling only, and is therefore somewhat speculative. Recreating similar conditions in the laboratory will unlock the physics involved in some of the most spectacular events in the known Universe, such as gamma-ray bursts and astrophysical jets.

\subsection{Phase-space control}\label{Sec:Phase-spacecontrol}
The development of techniques for controlling the transverse and longitudinal phase space of particle bunches in conventional accelerators took decades, and although many of these methods can also be used with plasma accelerators, the unique properties of the particle bunches they generate will require novel approaches to be developed. Many important applications of plasma accelerators require improvements in, or control of, the emittance, energy spread, or energy chirp of the particle bunch. The development of methods for phase-space control in plasma accelerators is therefore vital.

\begin{recommendation}
\label{Rec:ProgrammaticFunding}
Programmatic funding of plasma wakefield accelerators should be provided to optimise the quality and stability of plasma-accelerated particle bunches, to increase pulse repetition rates, and to enable key applications in the industrial sector and in fundamental science.
\end{recommendation}

\subsubsection{Low emittance}
The transverse emittance is a key parameter both for light source and high energy physics applications. The normalised emittance $\epsilon_n = \gamma \epsilon$, where $\epsilon$ is the geometrical emittance, is a crucial parameter for FELs since the lasing threshold diffraction limit for a FEL operating at a wavelength of $\lambda_\mathrm{rad}$ is $\epsilon = \epsilon_n/\gamma < \lambda_\mathrm{rad} /(4\pi) \propto \lambda_\mathrm{rad}^{1/2}$, where the proportionality holds for a given undulator.  Hence  FEL operation at shorter wavelengths requires decreased normalised emittance \cite{LiAdelmannXFEL2005}, which becomes increasingly challenging . The normalised emittance also determines the beam brightness  $B \propto \epsilon_n^2$, which in turn determines the FEL gain. For high energy physics (HEP) applications, the normalised emittance is a vital parameter since it determines the event rate: a low emittance enables a  final focus of small area, $\sigma_x\sigma_y$, and hence a high luminosity $ L = \frac{f N^2}{4 \pi \sigma_x\sigma_y} $, where $f$ is the frequency and $N^2$ is the number of particles.

Although measurement of emittance is difficult, especially for beams with significant shot-to-shot jitter and energy spread, there is solid evidence\cite{Brunetti:2010} that laser--plasma accelerators today routinely produce beams with $\epsilon_n \approx \unit[1]{mm\,mrad}$, depending on the electron injection method and laser--plasma parameters. \cite{PhysRevLett.119.104801BarberEmittance2017} These emittance values put LWFA-generated electron beams at the same level as those used in state-of-the-art X-ray FELs. Nevertheless, in view of the fundamental importance of emittance for key applications, and the potential for significant further improvements, substantial R\&D is required on methods to further improve and characterise the emittance. This includes the development of novel measurement techniques, since emittance is  a rather indirect observable and the measurement of low emittance beams is non-trivial. 

UK groups have played a leading role in developing new approaches which promise to decrease the obtainable emittance by three orders of magnitude  to the \unit{nm rad} level. These approaches, known as plasma photocathodes or Trojan Horse injection,\cite{Hidding:2012} are based on injection of electrons field-ionised from ions in the plasma by a trailing (injection) laser pulse. Achieving the lowest possible emittance requires that the intensity of the injection pulse is low, in order to minimise the transverse momentum gained by the ionised electrons, and that the electric fields of the  driver are lower than that of the injection pulse, to avoid ionization by the driver. These conditions favour PWFAs since particle beam drivers excite strong wakefields at electric field values of a few $\unit{GV m^{-1}}$  compared to the $\unit{TV\,m^{-1}}$  driver fields typical of LWFAs. The plasma photocathode approach has been recently realised for the first time in the E210:\,Trojan Horse PWFA experiment at SLAC FACET.

Plasma photocathodes could also be realised in LWFAs by using long-wavelength\cite{Yu:2014} or multi-pulse laser  drivers,\cite{Tomassini:2018vnw} and shorter wavelength injection pulses. These approaches deserve significant attention, including the development of high-power long wavelength laser pulses. In the UK, the high-power laser centres and CLARA offer promising facilities to develop this future class of plasma photocathodes.

\subsubsection{Low energy spread}
A key parameter in determining the quality of a relativistic electron beam is its spectral bandwidth, or relative energy spread. It is usually advantageous for the energy spread to be as small as possible. For example, a narrow energy spread is important for efficient acceleration in subsequent accelerator stages, and for subsequent beam manipulation and focusing. Narrowband electron beams are particularly important for driving light sources: for example, hard X-rays FELs require relative energy spreads  $\lesssim  0.1\%$, which at present can only be achieved for multi-GeV beams by conventional accelerators.

Both LWFAs and PWFAs can generate multi-GeV electron beams with relative energy spreads typically of a few percent \cite{Sasaki:2011, Leemans:2014lka,Litos2014Natureshort}. A  reduction in the energy spread by one or two orders magnitude is therefore required.  To a large degree the comparably large energy spread of plasma accelerators is caused by the high accelerating fields and small plasma cavity size; these features also cause the bunch to develop a negative longitudinal energy chirp  in addition to uncorrelated energy spread arising from the injection process. The total energy spread is usually dominated by the correlated energy spread, and this can be reduced  by various approaches including:  longitudinal phase space rotation in a beam transport section; periodically localizing the bunch at different phases of  the accelerating wakefield; or by beam loading.  These approaches are generally viable both for LWFAs and PWFAs, while in LWFAs dephasing is an additional complexity which needs to be taken into account. The compensation of local electric field gradient by matched beam loading or counter-rotation of longitudinal phase space by overloading the wake is directly connected to injection processes and control of these processes.

Reducing both correlated and  uncorrelated energy spread requires control of the injection of particles in the plasma wakefield. A wide variety of schemes has been proposed and are currently being studied experimentally and theoretically. A long-term scientific and technological development is certainly required in this area, including high-precision tailoring and stability of plasma sources, fine control of the spatio-temporal properties of the (laser or particle) driver beam, and the development of detailed control of the injection dynamics. Numerical and analytical work suggests that substantial improvements can be achieved in this area, down to relative energy spreads below 0.01\% for few-GeV electron beams \cite{NatCommmanahan2017}.

\subsubsection{High peak brightness}
\begin{figure}[tb]
    \centering
    \includegraphics[width = 100mm]{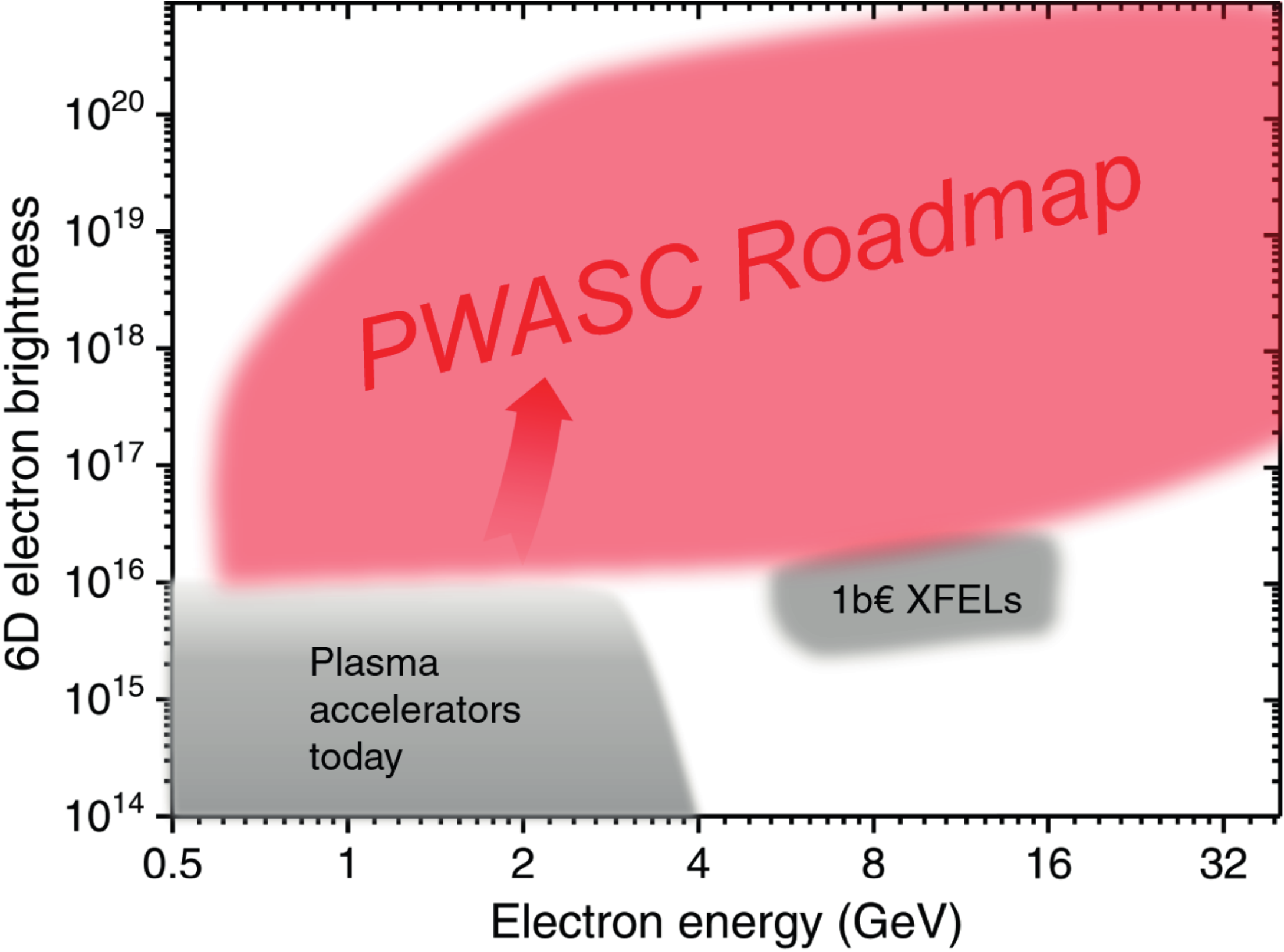}
    \caption{The 6D brightness of electron beams generated by current and future generations of plasma accelerators.}
    \label{Fig:6Dbrightness}
\end{figure}

The brightness of a particle beam \cite{Worster:1986} is a key performance parameter of accelerator output and particularly important for light sources \cite{photonics2020317dimitri}. It combines the particle beam current $I$ and the emittance $B_{\mathrm{5D}} = I/ \epsilon_n^2$ and in the so-called 6D-brightness also includes the relative energy spread. Plasma wakefield acceleration intrinsically generate ultrashort bunches with very high current, and can produce very low emittance values. Figure \ref{Fig:6Dbrightness} shows the 6D brightness of present and future generations of plasma accelerators as a culmination of prospects which are within reach of the approaches described in this roadmap. 

 At these extreme values, a notable challenge in particular for beams with significant energy spreads at the end of the plasma is extraction and transport of beams under preservation of beam quality, which can be addressed by beam manipulation via plasma-based devices \cite{vanTilborgJ:2015eo, Lindstrom:2018dr, Pompili:2018ths}. These plasma-based building blocks also allow the compactness and relatively low cost of the overall system to be preserved. 

High brightness is particularly important  for key applications, such as light sources \cite{photonics2020317dimitri}, as it may allow realization of high-gain coherent light sources with extreme performance in the hard x-ray range. Low emittance and energy spread are requirements shared with high energy physics applications of  ultra-relativistic particle beams in the form of beam luminosity, which is of fundamental importance for colliders.

As discussed in \S \ref{Sec:Next_gen_laser_drivers}, considerable effort world-wide is aimed at developing high-power laser systems capable of operation with repetition rates in the \unit[0.1 -- 10]{kHz} range, and we note that the development of \unit[1]{kHz} plasma accelerators is one of the main long-term scientific goals of the US accelerator roadmap. Such developments would be needed to enhance both the average brightness and luminosity of plasma-accelerated particle beams and  will dramatically boost the range of applications of laser-driven particle accelerators.

\subsubsection{Controlling injection}\label{Sec:Controlled_injection}
Since the accelerating structure typically travels at very close to the speed of light, in order to be trapped in a plasma accelerator, injected particles must rapidly achieve relativistic velocities. The simplest method of achieving injection is to drive the plasma wave to wave-breaking so that some background plasma electrons can enter the accelerating structure and become trapped. In LWFAs this mechanism has been widely studied for more than a decade and is responsible for the highest observed electron energies to date \cite{Leemans:2014lka, Gonsalves:2019ht}. However, this self-injection process is very sensitive to small variations of the laser--plasma parameters and is therefore usually associated with large shot-to-shot jitter of the bunch parameters.

Reducing shot-to-shot jitter, and improving or controlling the 6D phase space of the particle bunches, requires control of the particle injection process. 
A wide variety of techniques for controlling electron injection has been investigated to date. These include: locally reducing the wake velocity at density down-ramps;\cite{Geddes:2008} ionizing electrons from dopant species, by the driver or additional pulses;\cite{Hidding:2012, Yu:2014} localised stochastic heating with one or more colliding laser pulses.\cite{Faure:2006}

It should be noted that it is likely that no single method will be best suited for all applications. For example, radiation hardness testing requires high average charge, but beam quality is less important, whereas high bunch quality is vital for driving radiation sources.

Due to their experimental complexity, the more advanced injection mechanisms have been investigated less intensively, especially for the highest power laser systems. This area of research will become increasingly important as applications of plasma accelerators are developed, and progress will require dedicated beam time both to develop and optimise controlled injection techniques  and to make challenging measurements of the bunch quality.

\subsubsection{Tunability}
Plasma wakefield accelerators have the potential to offer wide tunability. For example, in a LWFA the particle energy can be tuned from keV to few GeV by adjusting the laser power and/or the length and density of the plasma; the bunch charge can vary from fC to nC, and consequently the current can be ramped up to the tens of kA scale. That said, the bunch parameters are not independent; for example, the peak energy and energy spread depend on the bunch charge via  beam loading.

\subsection{Transport, staging, and feedback}\label{Sec:Transport_staging_feedback}
\subsubsection{Beam capture \& transport}\label{Sec:Capture_and_Transport}

The successful incorporation of plasma accelerator concepts into scientific, engineering and industrial applications will in many cases require the controlled capture and transport of beams as they exit the plasma medium. Drawing on comparisons from advanced RF-driven particle accelerators, the transport challenges may be a limiting factor in the utilisation of the beams, as well as being an important factor determining the overall scale of any plasma accelerator facility. For example, the particle-beam transport optics and phase-space manipulation section in state-of-the-art FELs and high-energy physics machines \cite{CLIC-CDR} constitute a substantial part (and cost)  of the overall system footprint. 

Plasma wakefield accelerated  beams present additional challenges  beyond those of RF-driven accelerators. The sub-micron transverse dimensions together with the very high  divergence on exiting the plasma medium require focusing magnetic optics with strengths beyond the magnetic saturation levels of Fe-based electromagnets. Ceramic permanent magnets may be a solution since they provide high field gradients, but further development is needed to meet the field-quality levels, such as field flatness, required for many applications. Other approaches, such as active plasma lenses are being explored by UK and international groups.\cite{vanTilborgJ:2015eo, Lindstrom:2018dr, Pompili:2018ths}. More exotic concepts such as large energy acceptance `neutrino horns` may also offer innovative solutions. 

At few-percent energy spread levels, much tighter  control of chromatic aberrations than generally encountered in accelerator applications is required. Corresponding setups  may require  sextupoles supplemented with octopole and higher order magnets. Minimization of the energy spread before extraction from the plasma stage is therefore crucial to reduce the complexity and challenges of beam transport.  

The intrinsically ultrashort duration and high current of plasma generated beams is a highly attractive feature, but also a substantial challenge to beam control. The associated high current density is known to cause significant coherent synchrotron radiation (CSR) during transport through dispersive elements, which causes back-reactions in the particle beam phase-space. Related deleterious processes such as the excitation of the CSR microbunching instability may occur, although the growth of such instabilities may be damped by the large energy spread of the beams; successful efforts to reduce the energy spread for applications will give rise to increased need to address coherent and collective instabilities in the transport of short duration high-current beams.

Some applications may benefit from non-conventional transport optics. An example occurs in efforts to obtain coherent gain and free-electron laser action; lasing requires very small uncorrelated energy spreads, to limit Landau damping in the microbunching formation. This stringent constraint can however be mitigated though a combination of transverse dispersion of the beams, coupled with transverse gradient undulators. Such concepts are currently being pursued experimentally by several groups internationally, and are likely to be of interest in UK efforts in FEL action with PWFA beams.

\subsubsection{Staging}
Laser-driven plasma accelerators have demonstrated multi-GeV energy gain in a single accelerator stage (i.e.\ a single target, a single driver beam). Although it will be possible to extend beyond the \unit[10]{GeV} range with planned facilities (such as ELI, Apollon, CLF \unit[20]{PW}), large laser installations of this kind are currently limited to low repetition rates, which limits the potential applications and makes it difficult to maintain long-term accelerator stability.

An alternative approach to obtaining higher beam energy is staging of multiple plasma accelerator stages, each driven by its own laser pump. This rephases the accelerating structure with the accelerated beam, eliminating the effects of dephasing or pump energy depletion. In principle, the particle beam energy can then be increased limitlessly by increasing the number of stages. Staging presents three main challenges: (i) transporting the particle beam between stages; (ii) coupling in the driver beam into each stage; and (iii) stability. 

As discussed in \S \ref{Sec:Capture_and_Transport}, transporting the particle beam between stages necessarily increases the total length of the accelerator and novel methods need to be developed to deal with the challenging parameters of plasma-accelerated beams. To meet this challenge, active areas of research include the development of active plasma lenses \cite{vanTilborgJ:2015eo, Lindstrom:2018dr, Pompili:2018ths},  and shaping of the plasma to directly couple beams from one stage to another. 

Coupling the laser pulse into each stage requires an optic of several metres focal length which would decrease the average acceleration gradient substantially if the drive focusing and accelerator stages were arranged in a linear geometry. A much better option would be to fold the drive beamline so that most of its length is perpendicular to the axis of the accelerator, and only the last few centimetres are directed along the accelerator axis by a mirror. This mirror must withstand high laser intensities and allow transmission of the particle beam from the previous stage. A promising solution is a plasma mirror: this comprises a thin tape which allows transmission of the particle beam with minimal emittance growth; the incident laser pulse converts the surface of the tape into a dense plasma, and hence is reflected with high efficiency. Plasma mirrors have been used to couple two plasma stages to yield electron beams in the \unit[100]{MeV} range, \cite{Steinke:2016ga}  although implementation in the GeV-range with higher efficiency has yet to be achieved. Innovative solutions such as curved plasma channels \cite{PhysRevLett.Bentstaging2018} for staging are therefore also being developed.

The small size of the wakefield structure means that coupling driver and particle beams into each stage requires achieveing, and maintaiting, alignment to the micron level or even better. Achieving this will require active pointing stabilization of both the laser system and the plasma stages.

Although dephasing is not a relevant limitation in  PWFAs and tens of GeV energies have been realised in single stages, staging is required if aiming at electron energies which exceed low multiples of the driver beam energy, determined by the transformer ratio. Although conventional transport systems could be used to couple driver and witness beams into each stage, the distances required will reduce the mean accelerating gradient. Novel approaches, such as off-axis injection, are therefore being investigated to maintain a high average acceleration gradient.

One of the principal motivations for proton-driven PWFAs is the potential to accelerate electrons from the GeV to the TeV scale in one stage due to the high stored energy of the drive bunches in the CERN SPS or LHC accelerators.  In this this way, it is possible to avoid the issues of staging.

\subsubsection{Advanced diagnostics}
The characterisation of particle beam phase space is an important aspect for developing acceleration processes, benchmarking with simulation codes, and for  exploitation of plasma accelerated beams for applications. The accessible  parameter regime goes well beyond that known from  RF-accelerated beams; or where measurement techniques exist in the conventional accelerator realm, the techniques may not be applicable to the facilities and environment of PWFA experiments. The measurement of particle phase-space with femtosecond time resolution or better, and ultralow emittance, are particular challenges. In RF-driven accelerator facilities temporal characterisation is achieved with RF powered transverse deflection structures (TDS) providing a temporal streak to the beam. TDSs are capable of measuring the slice-parameters of energy vs time, or emittance vs time, with few femtosecond resolution. The measurement, and the time resolution, are however entwined with beam transport, and typically require tens of metres transport and magnetic focusing and defocusing optics. The RF infrastructure requirements are large, in the region of a few million pounds. To date TDS measurements of LWFA beams have not been demonstrated. Beam-driven plasma accelerators are more amenable to incorporating such diagnostics.

Coherent optical emission has been more widely explored for temporal measurements of charge density. While ambiguities exist in inferring temporal information from optical spectral emission, these techniques have demonstrated capability in determining plasma acceleration physics, e.g. through observation of multiple-bucket acceleration. Coherent emission techniques are not currently amenable to slice-parameter characterisation.

There is a need for new concepts and techniques for accessing temporal phase-space information. Candidates include THz-driven streaking of electron beams, which are being developed for RF-driven beams, and in the streaking of X-ray-liberated photoelectrons for fs X-rays diagnostics, or by exploitation of the plasma afterglow response. For a sufficiently large project directed at plasma acceleration there is potential to include a RF driven TDS diagnostic for plasma accelerated beams. 

There is also a need for new concepts and techniques for transverse phase space measurements to resolve the nm rad-scale normalised emittance values, and corresponding brightnesses, which may be achievable by novel bunch generation methods as outlined in \S \ref{Sec:Phase-spacecontrol}.

\subsubsection{Stability, feedback and control}
\subsubsection*{Stability}\label{Sec:Stability}
At present laser--plasma accelerators typically exhibit relatively large shot-to-shot jitter of the bunch parameters, including energy, charge and  energy spread. For example, shot-to-shot variations in these bunch parameters often range up to tens of \% and typically also are combined with pointing variations which are much higher than those of conventional accelerators.

There are two reasons for the relatively poor stability typically demonstrated to date for laser-plasma accelerators: (i) in most work to date, electrons are self-injected into the wakefield by stochastic processes which are therefore prone to large shot-to-shot variations; (ii) the parameters of the driving laser, and those of the plasma target to some extent, themselves show large jitter and drift. As discussed in \S \ref{Sec:Controlled_injection}, the solution to (i) is to develop methods for controlling the injection (and trapping) of electrons into the wakefield, and ideally to decouple it from driver and plasma shot-to-shot variations as far as possible. The solution to (ii) includes general advances in driver pulse technology and to develop feedback and control systems.

We note that the electron and proton drive bunches used in PWFA have well-defined energies, with little variation and a small energy spread. These positive features will help ensure that the shot-to-shot stability of PWFAs is high, and could be improved further by incorporating feedback systems.

\subsubsection*{Feedback and control}\label{Sec:Feedback_control}
Lasers capable of driving GeV class plasma wakefield accelerators are now able to operate at repetition rates at \unit[1]{Hz}, and are expected to develop to $> \unit[10]{Hz}$ in upcoming facilities. The increased repetition rate will allow for enhanced active feedback on laser and target parameters for the optimisation and stabilisation of LWFAs. 

There are many causes of drift and shot-to-shot jitter in the bunch parameters of a LWFA including, diurnal variation of the environment, slow heating of the laser systems with use, vibration, powerline fluctuations, and drift or fluctuations in the density of the plasma source. The stability of LWFAs against  jitter and drift could be improved significantly by implementing feedback systems for the drive laser and plasma stages. Control systems could include feedback loops for the laser energy, beam position, and beam pointing, together with active control of the position and density of the plasma stage. Control and feedback systems of this type would be relatively straightforward to implement, but to date they have not been deployed at national laser facilities (such as CLF), since these are general purpose facilities which need to cater to a wide range of experiments.

At a higher-order level, feedback and control could be used to adjust the laser and plasma parameters pro-actively to stabilise the output bunch parameters. This requires knowledge of the matrix connecting the output bunch parameters with the laser and plasma parameters.

Feedback loops can also be used to \emph{optimise} the output beam parameters. Proof-of-principle experiments have already demonstrated the use of this approach to optimise the charge and divergence of sub-MeV electron beams at \unit[1]{kHz} \cite{He:2015}, and of \unit[$\sim 100$]{MeV} beams at \unit[5]{Hz}.

We note that PWFAs could operate at repetition rates up to the MHz range. The high repetition rate is expected to make it easier to develop feedback and control systems, and these can draw on decades of development for conventional accelerators. The development of high repetition rate feedback and control systems for PWFAs will be beneficial for all plasma wakefield research.

% Synergies and impact
\section{Applications \& impact of plasma accelerators}\label{Sec:Applications}
\begin{figure}[tb]
\centering
\includegraphics[width = 0.75 \linewidth]{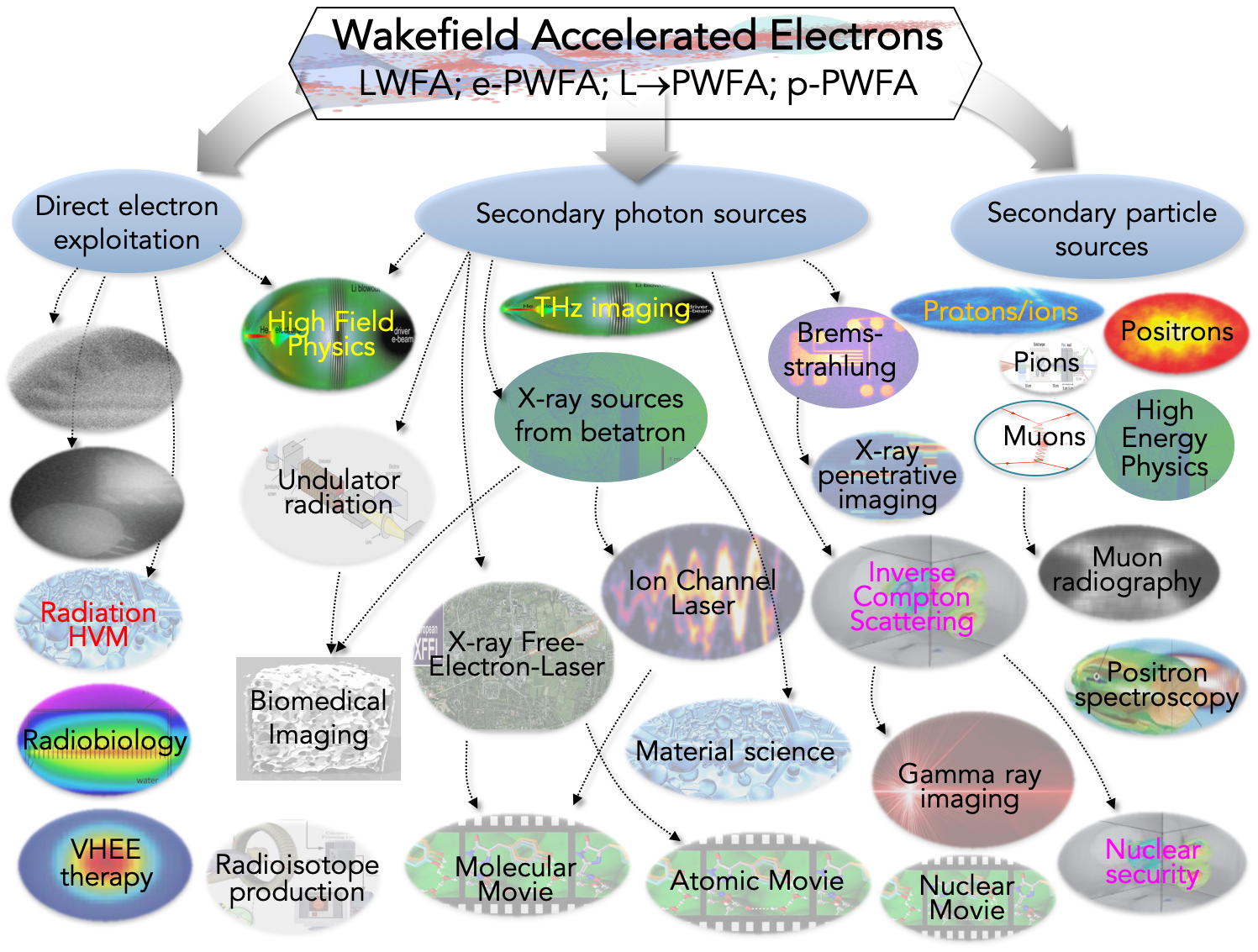}
\caption{Schematic diagram showing a subset of  potential applications of plasma wakefield accelerators, with impact across various areas of science and industry.}
\label{Fig:PWASC_Multi-council_impact}
\end{figure}

Plasma accelerators have the potential to be a disruptive technology which could fundamentally impact many areas of science, technology and medicine, as illustrated in Fig.\ \ref{Fig:PWASC_Multi-council_impact}.

In order for plasma accelerators to be developed and deliver solutions for industry at the appropriate Technology Readiness Level (TRL), industrial representatives are needed to champion the emerging disruptive technologies and to provide the industry pull and guidance.  We have identified collaborative opportunities in the space technology area, energy, aerospace, nuclear, advanced manufacturing, new materials, security and defence sector and medicine. Some industrial links ranging from SME's to major players have already been established. In view of the greater appetite for industry collaboration, and increased industry interest as the area matures, it is recognised that challenges led by industry consortia are of great importance.

The plasma wakefield community will benefit from strong links and knowledge transfer with other areas of academia to influence and inspire future science and development areas. Academic groups such as accelerator science, particle physics, medical/radiobiology and materials engineering are identified as core opportunities for interdisciplinary research activity and collaboration. Cross-council funding mechanisms which can support these important activities are urgently needed. The emergence of the Industrial Strategy Challenge Fund (ISCF) is  welcome, although the community has had very limited influence on this so far. Owing to the fundamental character of plasma accelerator research, and the wide TRL range of their various applications, a targeted sector initiative may be required.   

\begin{recommendation}
\label{Rec:Cross-Council}
UKRI should develop mechanisms for providing cross-council support for the wide range of research, in a variety of settings,  necessary to drive advances in plasma accelerators; this range includes fundamental research (e.g.\ plasma physics), technology development (e.g.\ novel lasers), and application development (e.g.\ medical imaging).
\end{recommendation}

\subsection{Radiation sources}
A near term application of plasma accelerators is driving compact radiation sources. Undulator radiation with photon energies of about \unit[100]{eV} has been generated from laser-accelerated electrons,\cite{Fuchs:2009} and the betatron motion of the particle bunch has already been used to generate incoherent X-rays in the \unit[10]{keV} range,\cite{Kneip:2010} which can be increased beyond \unit[100]{keV} by driving the oscillation of the electron bunch with a laser pulse.\cite{Cipiccia:2011} Incoherent Thomson- or Compton-scattered radiation in the \unit[10]{keV} to \unit[1]{MeV} range \cite{Phuoc:2012vb, Powers:2013bx} has been generated by colliding plasma-accelerated electron beams with an intense, counter-propagating laser pulse.

An attractive feature of radiation sources driven by plasma accelerators is that the short duration of the electron bunches leads directly to the generation of femtosecond-duration radiation pulses. For LWFAs the radiation will also be naturally synchronised to an ultrafast laser system able to generate auxiliary pump or probe pulses, thereby providing an unprecedented and versatile tool for ultrafast science. Radiation sources driven by plasma accelerators could be sufficiently compact and cheap to bring  ultrafast imaging and probing techniques from large-scale facilities into small-scale labs, thereby making them ubiquitous and opening new horizons in many areas of the medical, biological and physical sciences.

Plasma-accelerator-driven radiation sources are now starting to be used in applications. For example, betatron X-rays have been used to produce tomographic images of biological samples (see Fig.\ \ref{Fig:Medical_imaging}),\cite{Cole:2015fh} and to capture the propagation of shock fronts in laser-shocked materials.

To date, radiation sources driven by plasma accelerators have been incoherent. A medium term goal is driving compact FELs with a peak spectral brightness ten orders of magnitude brighter than incoherent sources. X-ray FELs (XFELs) driven by kilometre-scale conventional accelerators have transformed many areas of ultrafast science and are producing a torrent of high-impact results across the physical and life sciences. However, their large scale, and high cost (of order \pounds1 billion) means that  only a handful of such facilities exist worldwide. The impact of developing XFELs driven by plasma accelerators is self-evident. The challenge to be overcome is the generation of electron bunches of sufficiently high quality for free-electron gain to occur at short wavelengths. This problem is being pursued very actively, particularly in Europe, Japan, and the USA, and it seems likely that the first gain demonstration of a plasma-accelerator-driven FEL can be achieved in the next five years. The first demonstrations will probably be at relatively low photon energies, but further improvements in the quality of the bunches generated by plasma accelerators will lead to FEL operation into the X-ray spectral range.

We note that future advanced radiation sources driven by \emph{conventional} accelerators are likely to employ ultra-short electron bunches, which in turn will require the development of new diagnostics for characterizing electron bunches of ultra-short duration and low emittance. The bunches generated by plasma accelerators could provide an ideal source for testing and characterizing the new diagnostics which will be required. Diagnostics are therefore another area, in addition to driver technology, where conventional and novel plasma accelerator mechanisms merge and cross-fertilise each other.

\subsection{Medical applications}
Plasma-based accelerators, especially those driven by laser fields have the key advantage of being able to drive multiple bright sources of energetic particles and high energy X-rays beams from a single machine. With appropriate optimisation, these beams have the potential to provide a unique capability for biomedical imaging, cancer diagnosis and therapy from a single facility. 

\subsubsection{High-resolution X-rays sources for biomedical imaging}
Early detection of cancer is one of the crucial factors determining the probability of surviving it. Existing screening methods such as digital mammography and CT have poor discrimination between glandular and tumour tissues because of their similar X-ray attenuation. A new imaging method sensitive to the phase of X-rays, rather than just their absorption, yields enhanced intra-tumour soft-tissue contrast and improved visualization of cancerous structures, especially in soft tissues,  and can potentially be achieved with a lower dose to the patient.  However, improved patient outcome so far has only been demonstrated at large and expensive synchrotron sources, which have limited access for medical use. Compact betatron radiation sources driven by plasma accelerators could make earlier diagnosis routinely available, transforming treatment planning, delivery, and monitoring.

\begin{figure}
\centering
\subfloat[][]{\includegraphics[height = 60mm]{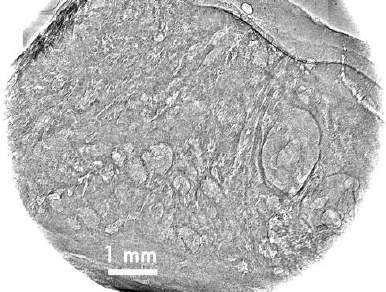}}
\hspace{2cm}
\subfloat[][]{\includegraphics[height = 60mm]{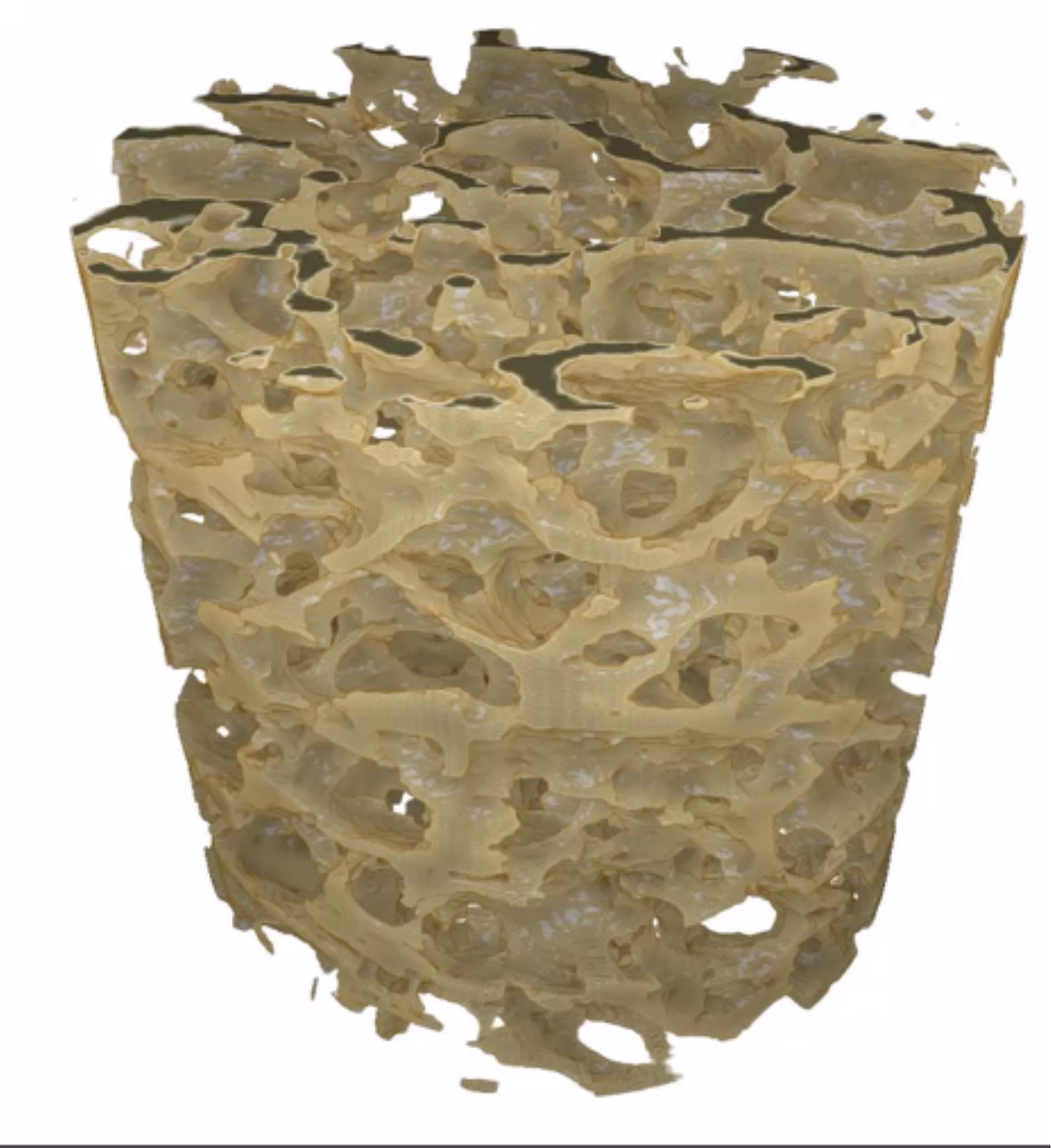}}
\caption{Medical imaging with laser-driven plasma accelerators. (a) Phase contrast image of prostate tissue; (b) tomographic image of a human trabecular bone using laser-driven accelerators.}
\label{Fig:Medical_imaging}
\end{figure}

The high flux of LWFA-driven betatron sources means that images can be acquired in a single laser pulse, overcoming the disadvantage of a micro-CT system which require a long exposure time for a high-resolution tomographic scan. Figure \ref{Fig:Medical_imaging} shows the results of phase contrast and tomographic imaging of human tissue samples obtained at CLF by UK groups.  The small size of the laser-based source makes them suitable for deployment in a hospital environment, which is much preferred for clinical applications.

\subsubsection{Very high-energy electron beams for therapy (VHEEs)}
A new radiotherapy (RT) modality uses so-called very high-energy electrons (VHEEs, electron energy \unit[]>100MeV) to provide an alternative  treatment than X-rays for deep-seated tumours; this approach exploits the large momentum of VHEEs. Experiments and simulations have shown that VHEEs can have improved dose distributions over X-rays and can offer enhanced relative biological effectiveness (RBE). Conformal irradiation makes VHEEs ideal for targeting radio-resistant tumours. However, few VHEE-RT experimental studies have been undertaken because conventional accelerators are expensive. 

Laser-based VHEE radiotherapy is potentially disruptive technology that could cut the cost of cancer treatment. Initial studies performed at ALPHA-X/SCAPA indicate the suitability of VHEE for treating deep-seated tumours \cite{Moskvin:2012}. These studies have also shown that VHEE beams have similar biological effects to X-rays but that they have very different dose rate and dose distribution.

\subsection{Industrial applications}\label{Industrial_applications}
Industrial applications are an acknowledged strength of plasma accelerators\cite{web:STFCTech-X, web:EUCARD2ApplicationsAcceleratorsEurope, web:TIARA}. The flexibility of plasma accelerators and the ability to produce electron, proton, ion and photon beams, combined with their capability to be realised in ``table-top'' (or even portable) setups, makes them ideal instruments for industrial exploitation and for promoting transformative applications. The research and development of plasma accelerators meets most of the ten Industrial Strategy Pillars:

\begin{enumerate}
    \item[Pillar 1:] ``Investing in science, research and innovation'' is met by all plasma accelerator activities.
    
    \item[Pillar 2:] ``Developing skills'' is promoted by the interdisciplinary (lasers, beams, plasmas, high-end electronics) nature of novel accelerator science \& engineering training, e.g.\ at the accelerator institutes and universities.
    
    \item[Pillar 3:] ``Upgrading infrastructure'' is a central need for the plasma wakefield accelerator community (see \S \ref{Sec:Facilities}).
    
    \item[Pillar 4:] ``Supporting businesses to start and grow'' is achieved through the technology and innovation centres and incubators e.g.\ at Daresbury, Harwell, and at universities.
    
    \item[Pillar 5:] ``Improving procurement'' is driven by the need, for example, for highly specialised, high performance electronics, nano-fabricated structures, and laser systems for accelerator applications; this demand can provide commercialization opportunities which can put companies in a leading position before the need for mass production.
    
    \item[Pillar 8:] ``Cultivating world-leading sectors'' is an opportunity to build on the world-leading, and highly respected contribution to plasma accelerator research by the UK groups.
    
    \item[Pillar 9:] ``Driving growth across the whole country'' is promoted by a distributed novel accelerator infrastructure which encompasses university groups and application-oriented centres distributed across the country. The latter include: CALTA (Harwell), SCAPA (Glasgow), and VELA/CLARA (Daresbury). The expansion of these research programmes and facilities would foster growth in industrial development, with a high return on investment \cite{web:OECDInfrastructures}.
    
    \item[Pillar 10:] ``Creating the right institutions to bring together sectors and places'' can be achieved via a distributed accelerator infrastructure which builds on existing links between between national laboratories and facilities, universities and industry.
\end{enumerate} 

\begin{figure}
\centering
\includegraphics[width = \linewidth]{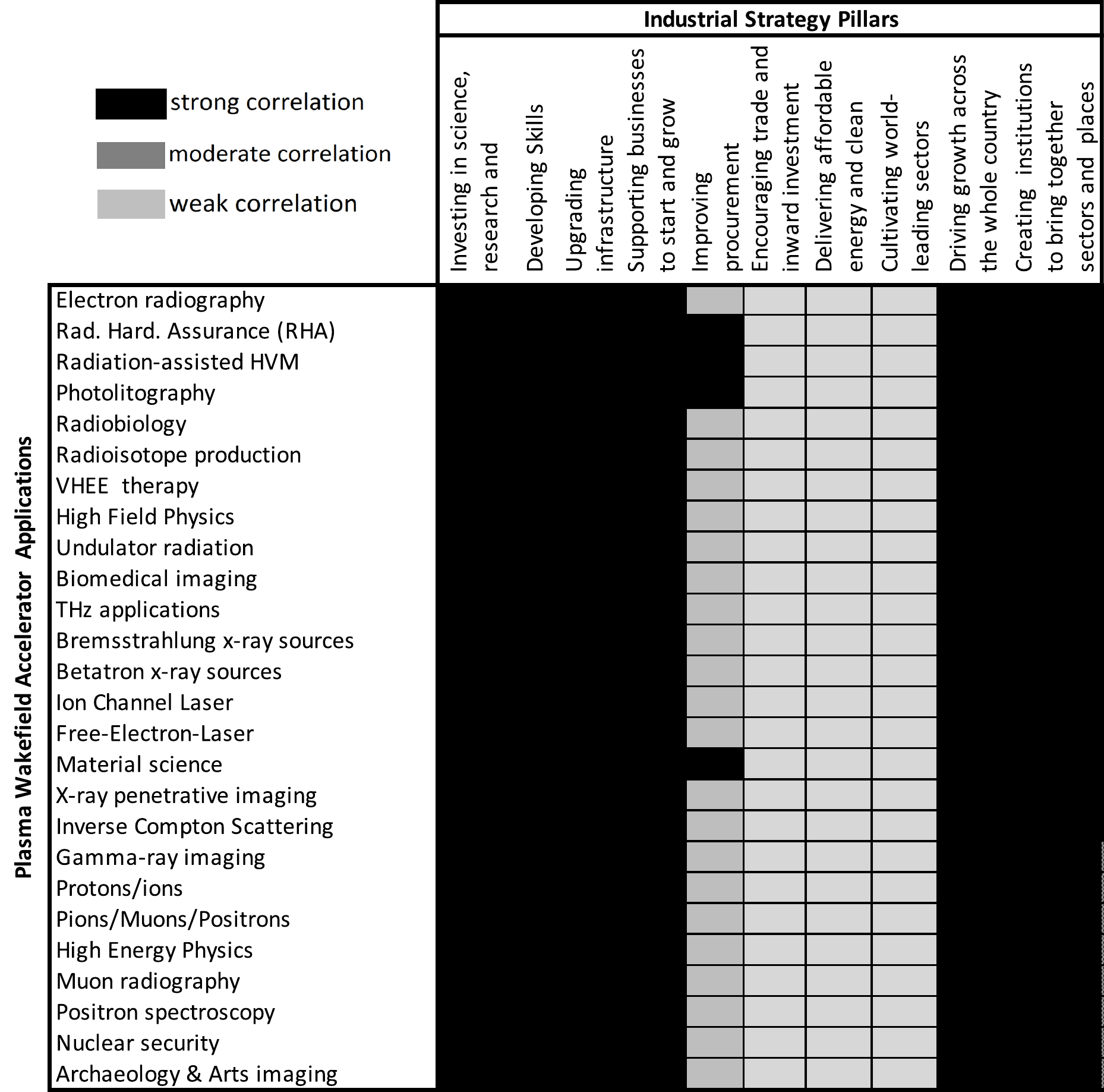}
\caption{Connections between applications of plasma wakefield accelerators and the 10 Industrial Strategy Pillars.}
\label{Fig:ISCF_map}
\end{figure}

Figure \ref{Fig:ISCF_map} maps selected plasma accelerator  applications to the 10 Industrial Strategy Pillars.

\subsection{Material science}
The properties and behaviour of materials depends on their structure at atomic and  macroscopic scales. The study of material properties at these scales is therefore important for understanding fundamental physics as well for developing applications in industry. Materials science is an exciting potential application for plasma-accelerator-driven radiation sources since they  can be used to probe the composition, structure and state of materials during processing or stress testing.

The industrial sector has shown considerable interest in  absorption and phase-contrast imaging measurements with LWFA-driven X-ray beams, \cite{Cole:2015fh,Wenz2015NC} and this is seen as a near-term opportunity for societal impact. The potential of LWFA-driven betatron sources to produce hard X-ray radiation over a wide bandwidth  allows for single-shot measurements of X-ray absorption spectra, thereby probing the atomic state of materials. The  short pulse duration of these sources, and their tight synchronization to a femtosecond laser system, offers opportunities for pump-probe experiments with unprecedented temporal precision. Indeed, penetrative imaging and X-ray spectroscopy (e.g.\ XANES or NEXAFS) with femtosecond resolution is beyond even the capabilities of state-of-the-art conventional light sources and is therefore an exciting near term application. Extension to higher photon energies, for example via bremsstrahlung or inverse Compton scattering (ICS), also provides the possibility of probing inner-atomic transitions for penetrative imaging of bulk ($\unit[10+]{cm}$ thick) materials.

% diffraction
With further optimisation of the spectral brightness of plasma-accelerator-driven photon sources it will be possible to perform X-ray diffraction experiments with crystalline and poly-crystalline samples. Determining the lattice structure, especially in pump probe experiments, is of great importance for understanding and developing novel materials, such as topological insulators \cite{HsiehN2008}. This work would require X-ray optics to transport the beam to the sample. An alternative approach would be to use  diffraction of electrons with energies in the \unit[10s --100s]{MeV} range, which could potentially be driven by lower energy, high repetition-rate laser systems. In this case, high charge, low energy spread, low emittance electron beams will be required, with suitable electron beam transport optics; as for the case of X-ray diffraction, the temporal resolution could be in the femtosecond range.

Both LWFAs and PWFAs can be purposefully used to generate broadband, exponential/power-law spectra particle beams to reproduce the spectral features of space radiation. This approach can be used for advanced radiation hardness assurance of space electronics and for nuclear environments \cite{srephidding}.

The key technical requirements for realising the near- and medium-term opportunities in materials science could be met by the development of dedicated plasma accelerator based X-ray/electron beamlines.

%% RECOMMENDATIONS

\subsection{Security applications}
For the security and defence sector, LWFAs  provides bright, highly penetrating sources of X-rays, electrons and neutrons for non-destructive evaluation (NDE), radiation hardness and damage testing and directed energy applications. A long standing programme of collaboration between the CLF user community and DSTL is testament to the interest from this sector and the potential for impact of laser accelerators in this area. 

Laser-driven X-ray radar has attracted great interest for security applications. This patented technique is enabled by the ultrashort pulse length and highly penetration capability of LWFA electrons and has been applied for depth profile or ``through barrier single-sided'' imaging. In this application, an important feature of laser-driven sources is the ability to  switch easily between different modes of radiation and particle beam generation, thus providing a highly versatile system with the potential for fast data acquisition. High resolution radiography is of great interest in this sector, from phase contrast imaging of composite materials used in armour to radiography of large components. The short pulse duration and high pulse brightness of LWFA-driven sources means that single shot exposure combined with high resolution imaging of objects is possible, which is highly valuable for material damage studies.  Laser wakefield beams are also being developed to generate MeV energy bursts of X-rays for interrogation of sealed containers to give element identification and to provide a method for rapid identification of special nuclear materials.

\subsection{High-energy physics}
Historically, the primary motivation for plasma wakefield acceleration has been its application to an electron-positron linear collider with energies in the TeV scale.  Given the size of conventional RF accelerators operating in this energy range (\unit[30 -- 50]{km}), the desire to significantly reduce the size and cost of particle colliders is strong.  This is backed up by the large energy gains observed in LWFAs and PWFAs, indicating that the accelerating section can be an order of magnitude (or more) shorter.

However, given that many cross sections fall with increasing energy and exotic physics beyond the Standard Model of particle physics is also expected to have a low production rate, colliders generally need to have high luminosity as well as high energy.  Building a plasma wakefield collider would therefore be challenging since it would require: a beam energy of hundreds of GeV, with an energy spread of 0.1\%; repetition rates of $>\unit[10]{kHz}$; bunches with $10^{10}$ particles and of nanometre transverse extent; and the generation of positrons with the same characteristics.  Further, the production of such beams needs to be reliable and reproducible so that a high luminosity, as well as high energy, can be achieved. 

Achieving high luminosity is a challenge for plasma wakefield acceleration and given its current status, there has not been a great deal of discussion amongst the HEP community on the possible applications. The development of particle colliders driven by plasma accelerators is therefore a goal which must be considered to be a long-term objective. A significant effort over an extended period, with extensive international collaboration, will be required to meet these considerable challenges. Several international programmes are coordinating efforts in this area, such as the AWAKE, Helmholtz Virtual Institute for plasma wakefield acceleration, and EuPRAXIA projects; the ICFA panel on Advanced and Novel Accelerators (ANA) has hosted several workshops on developing roadmaps for plasma accelerator colliders and has formed the Advanced LinEar collider study GROup (ALEGRO), to co-ordinate the preparation of a proposal for an advanced linear collider in the multi-TeV energy range. The UK has strong representation in all these efforts, and in order to exploit its strength in this area it will be important that the UK groups are able to continue to contribute to these and other collaborative efforts.  The US LWFA and PWFA communities have produced a similar roadmap with focus on developing both of these technologies towards a high-energy, high-luminosity linear $e^+e^-$ collider with the goal of technical design reports by the mid-2030s.  

Although, having an electron--positron linear collider as an ultimate aim for plasma wakefield acceleration and working towards achieving a number of its parameters is very valuable, it is prudent to consider other first applications. One should however distinguish between a stand-alone all plasma acceleration collider, and an upgrade of conventional collider with plasma acceleration.  While it is at this moment inconceivable to suggest an all-plasma $e^+e^-$ collider, it is reasonable to consider plasma acceleration upgrades for either the ILC or CLIC colliders if construction of the first Higgs-factory or top-factory stage of either is approved. Given the rate of the progress of plasma acceleration technology, it is entirely possible to consider TeV upgrades based on plasma acceleration.

\subsubsection{Electron--proton collisions}
Experiments in which high-energy electrons hit a target give insight into the fundamental structure of matter.  The HERA ep collider had a centre of mass energy of \unit[320]{GeV} and was the only such collider; all other experiments had a fixed target.  With a high energy electron beam ($>\unit[50]{GeV}$), further fixed-target experiments could be performed.  As many have been performed previously, one would need to take advantage of improved detector instrumentation and consider kinematic regions or physics still poorly understood, e.g.\ measurements which would have a strong impact for LHC physics.  This needs study and development: consideration of beam structure, review of past experiments, detector requirements and physics goals.

The Large Hadron electron Collider (LHeC) is a proposed ep collider using the \unit[7]{TeV} proton beam and electrons of about \unit[60]{GeV}.  One could consider generating the \unit[60]{GeV} electron bunches by plasma wakefield acceleration.  Current designs~\cite{Xia:2014ida} of this would lead to a luminosity $10^3$ to $10^4$ lower than the conventional scheme.  One could therefore consider a low-luminosity LHeC, in particular if the LHeC does not go ahead, called Plasma Electron Proton and Ion Collider (PEPIC), but potentially at much lower cost, as well as trying to increase the luminosity.

A more radical idea is for a very high energy electron proton (VHEeP) collider with a centre of mass energy of \unit[9]{TeV}, but modest luminosity~\cite{Caldwell:2016cmw}.  This uses the \unit[7]{TeV} proton bunches to collide and to generate the wakefield to provide a \unit[3]{TeV} electron beam.  Higher luminosity is always valuable, but the extra energy reach opens up interesting particle physics avenues even with modest luminosity.  A kinematic region can be investigated where limited luminosity is needed and where QCD and the structure of matter is not at all understood.  A workshop looking further at the physics opportunities took place in June 2017~\cite{web:vheep}.

However, the physics potential of low luminosity HEP experiments needs to be considered along with the potential saving on accelerator size.

\subsubsection{Search for a hidden sector}
As well as fixed-target experiments to investigate the (QCD) structure of matter, these and beam-dump experiments can be used to look for exotic physics such as ``dark photons'' which could be an explanation for dark matter and a number of other puzzles in physics.  Dark photons are expected to weakly couple between the Standard Model and a new Hidden Sector.  An experiment at CERN, NA64, is currently looking for dark photons using a \unit[50-100]{GeV} electron beam.  If this beam energy could be achieved with an AWAKE scheme, a factor of 1000 more electrons on target could be produced, greatly increasing the sensitivity to dark photons.  This is an ideal type of experiment as a first application as there are less strict constraints on beam quality.

Some of the above are being considered as a part of a CERN study on ``Physics Beyond Colliders''~\cite{web:pbc-wing}, although other applications could exist.

\subsubsection{Use as test beams}
High energy electrons beams are valuable tools for characterisation of detectors for high energy physics as well as accelerator systems and diagnostics.  As there are so few around the world, having another facility could be of potential value. The ability to generate short bunches could be useful for testing modern HEP detectors, which often require picosecond timing.  Further, the high particle flux per bunch could enable testing of saturation effects in detectors and an assessment of detector resilience under conditions of high occupancy and particle pile-up. Plasma-accelerated pencil beams could also be used to test ultra-high granularity calorimeters. ultralow emittance beams could be used as test beams for emittance preservation in staged approaches. These areas could prove to be  near-term applications of plasma wakefield acceleration; indeed it is one of the potential applications being considered by the EuPRAXIA project.

\subsection{High field physics}
Uniquely, LWFAs combine ultra-relativistic electron beams with high-intensity electromagnetic fields. The generation of GeV-scale electron beams ($\gamma > 2000$), together with focused laser intensities of $\unit[10^{20} - 10^{21}]{ W\,cm^{-2}}$ results in electric fields in the rest frame of the electron which are a significant fraction of the critical field of electrodynamics, i.e.\ the  Schwinger field. Several fascinating phenomena are predicted to occur in this high-field regime including vacuum polarisation, pair production from a pure photon--photon collision, and radiation reaction. These phenomena are not readily accessible by  conventional accelerators and to date only one experiment at SLAC has been able to study electron--laser interactions at a significant fraction of the Schwinger field. Recent experiments with the Gemini laser, using laser-accelerated electron beams, have hinted at quantum effects.\cite{Cole:2018kp}

Accessing this regime will help understanding the largely unprobed non-linear aspects of quantum electrodynamics (QED), one of the most advanced theories in modern physics. This highly non-linear regime of QED is interesting not only in its own merit, but also for the far-reaching implications that it has in a wide range of physics, such as particle physics, plasma physics, astrophysics, and cosmology. For  example,  strong field QED is needed to explain the radiative properties of ultra-massive objects, and is now included in models of particle acceleration with the next generation of high-intensity lasers, which will provide peak powers approaching \unit[10]{ PW} and focal intensities exceeding $\unit[10^{23}]{W\,cm^{-2}}$. These lasers (ELI, Apollon, Vulcan 20PW, EXCELS...), which are expected to be fully operational within the next few years, will have a transformational effect in laser--plasma physics and laser-driven particle acceleration. 

There is still a substantial level of ambiguity in the theoretical models of laser--plasma interactions at these ultra-high intensities, where QED effects cannot be neglected. Moreover, one can envisage the possibility of exploiting laser-driven ultra-relativistic electron beams to generate, via bremsstrahlung, high energy photon beams. One can then imagine the construction of the first pure photon--photon colliders in which GeV-scale \emph{photons} are collided with visible photons, allowing the study of exotic phenomena such as pure photon conversion into matter and vacuum polarisation.

In order to extract meaningful data from experiments in this regime it will be necessary to increase the repetition rate of LWFAs, reduce the energy spread and divergence, and increase the bunch charge.

\section{Synergies and related R\&D}
Research on plasma wakefield acceleration has significant synergies with R\&D  in  novel acceleration, light sources,  high energy density, and high field physics. These research areas are connected on many levels:  to a substantial degree they use the same facilities and systems, and many groups engage in these topics in addition to their work on plasma accelerators. Hence, investment in the infrastructure for LWFA research would also significantly benefit research on laser-driven proton and ion acceleration, and in high energy density physics. Investment into electron-beam driven plasma wakefield acceleration would also support dielectric structure wakefield acceleration and applications; and co-located laser--plasma particle and light sources for manipulation and probing of matter benefit FEL-enabled  research, as seen for example at the Matter in Extreme Conditions (MEC) \cite{web:MEC}
station at LCLS, and as planned for the Helmholtz International Beamline for Extreme Fields (HiBEF). \cite{web:HiBEF}

\subsection{Cross-council relevance}\label{Sec:Cross-council_relevance}
   
The science underpinning plasma accelerators lies at the intersection of optics, lasers, plasmas, and accelerator science; and their  applications span the  physical, medical, material and life sciences as well as industrial and defence applications. It is no surprise, then, that funding this area has proved challenging for funding bodies worldwide. For example in the US, plasma wakefield acceleration is largely funded by DoE's High Energy Physics programme, which does not directly include  the development of novel light sources.  In the UK, plasma wakefield acceleration spans the EPSRC and STFC remits. A natural consequence of the cross-disciplinary nature of plasma accelerator research areas is that it does not lie at the centre of gravity of any individual research council's remit, which can make it more difficult to obtain funding in this research area.

\begin{figure}[tb]
\centering
\includegraphics[width= \linewidth]{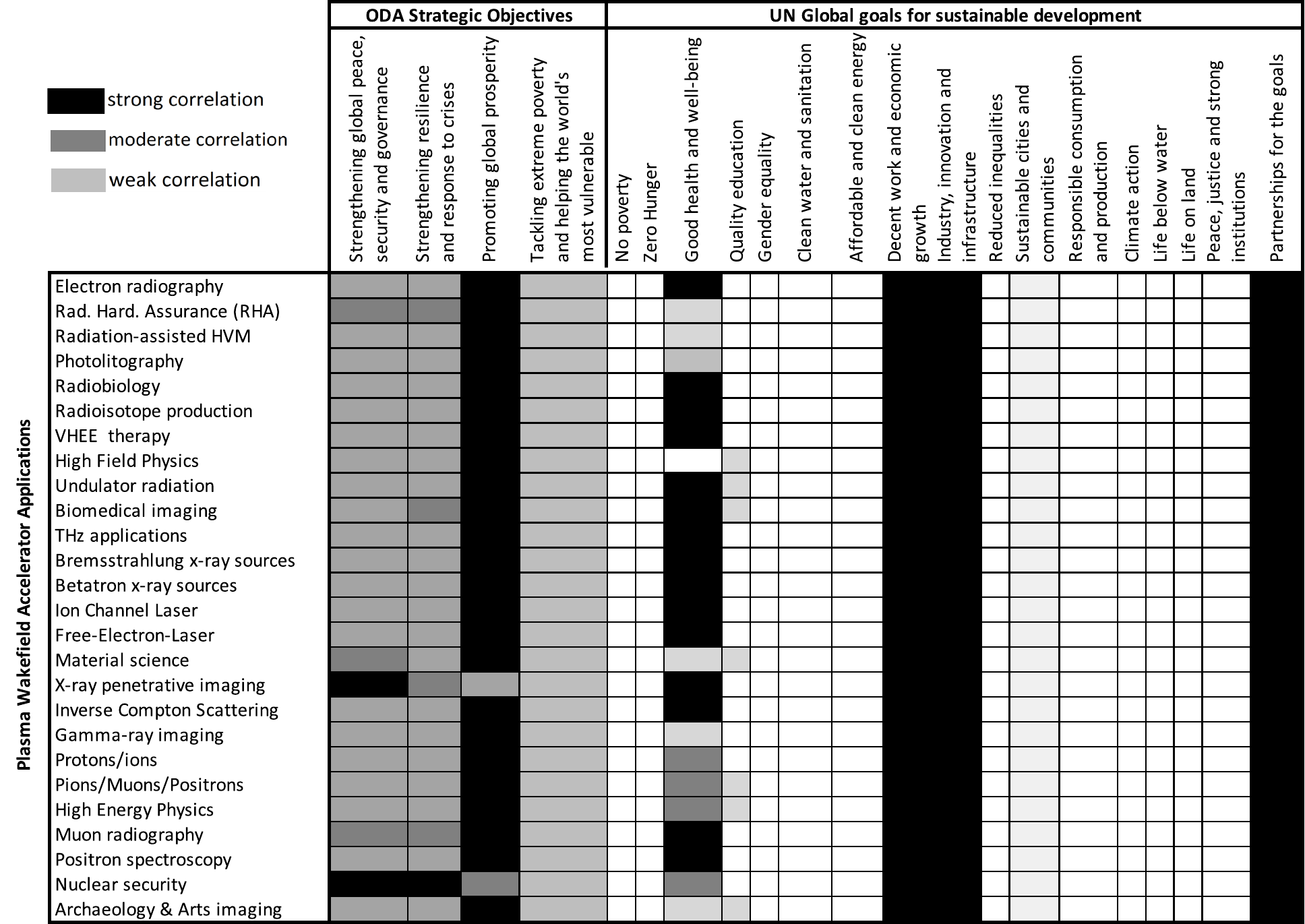}
\caption{Correlations between applications of plasma accelerators and the ODA Strategic Objectives and UN global goals.}
\label{Fig:ODA}
\end{figure}

The applications of plasma wakefield accelerators identified in \S \ref{Sec:Applications} illustrate the potential for commercial exploitation of opportunities which will emerge over the next several years. This will require investment in capabilities and capacities of UK facilities and laboratories, but also the development of funding mechanisms which can follow an application as it grows from concept to product. These opportunities match very well the increased UK emphasis  on industrialization of scientific research, such as the ISCF (see \S \ref{Industrial_applications} and Fig.\ \ref{Fig:ISCF_map}). 

The compactness, versatility and relative cost-effectiveness of plasma wakefield accelerators make them interesting for ODA activities in context of the GCRF. Figure \ref{Fig:ODA}, also used for the STFC2017/2018 Accelerator Strategy Review, shows that various established and increasingly maturing applications of plasma wakefield accelerators fit well to UK's ODA strategic objectives and UN global goals.

The need to foster multidisciplinary work, and to track potential applications from concept to product, would seem to make plasma wakefield accelerator research ideally suited to support through the recently established UKRI.  Various avenues may be envisaged to provide the UK plasma wakefield accelerator community with the increase in funding needed for them to exploit their internationally leading position. This could, for example, be achieved by an increase of the STFC Accelerator Programme budget, dedicated EPSRC funding programmes, and/or sustainable cross-council funding. In the recent STFC Accelerator Strategy Review Roadmap it was proposed to establish a cross-council Novel Accelerator Action Plan, which includes plasma accelerators, (non wakefield) plasma-based proton/ion acceleration, and dielectric and structure-based electron and laser-beam acceleration. This approach has been further substantiated by the accelerator community in the form of a selected potential large STFC ``priority project'' called UKNOVA.  

It is hoped that the present roadmap, together with community efforts such as the establishment of the PWASC, will help the UK Research Councils appreciate the potential impact of plasma wakefield accelerators, and to develop suitable mechanisms to enable the UK to maintain international leadership in this area.

% Timeline
\section{Timeline}
Plasma accelerators are experiencing worldwide growth, with many research projects and several facilities based on plasma accelerators and their applications being planned. Many of the breakthroughs achieved in this field have been led by UK research groups. However, as is often commented, the UK community has not always been successful to make the most of its own creativity, partly due to the difficulties of realizing appropriate funding levels discussed in \S \ref{Sec:Cross-council_relevance}. A major objective of this roadmap is to help ensure that the UK builds on its achievements to produce a new generation of dependable plasma accelerators for fundamental research and applications.

Although proof-of-principle experiments showcasing the capability of plasma accelerators  for several applications have already been performed, translating these into the real world requires improvements in beam stability and quality. The UK groups will focus on this in the immediate future. Since some of this work involves technology development rather than just high impact science, programmatic access slots are required in the national facilities such as at CLF. University labs and centres could also play an important role in this area. It is worth emphasizing that programmatic access will enable improvements to bunch quality and accelerator reliability which, although not obviously of the most immedidate scientific importance, will drive future scientific and technological advances with very high impact.  

\begin{landscape}
  \begin{figure}
      \centering
      \includegraphics[width=220mm]{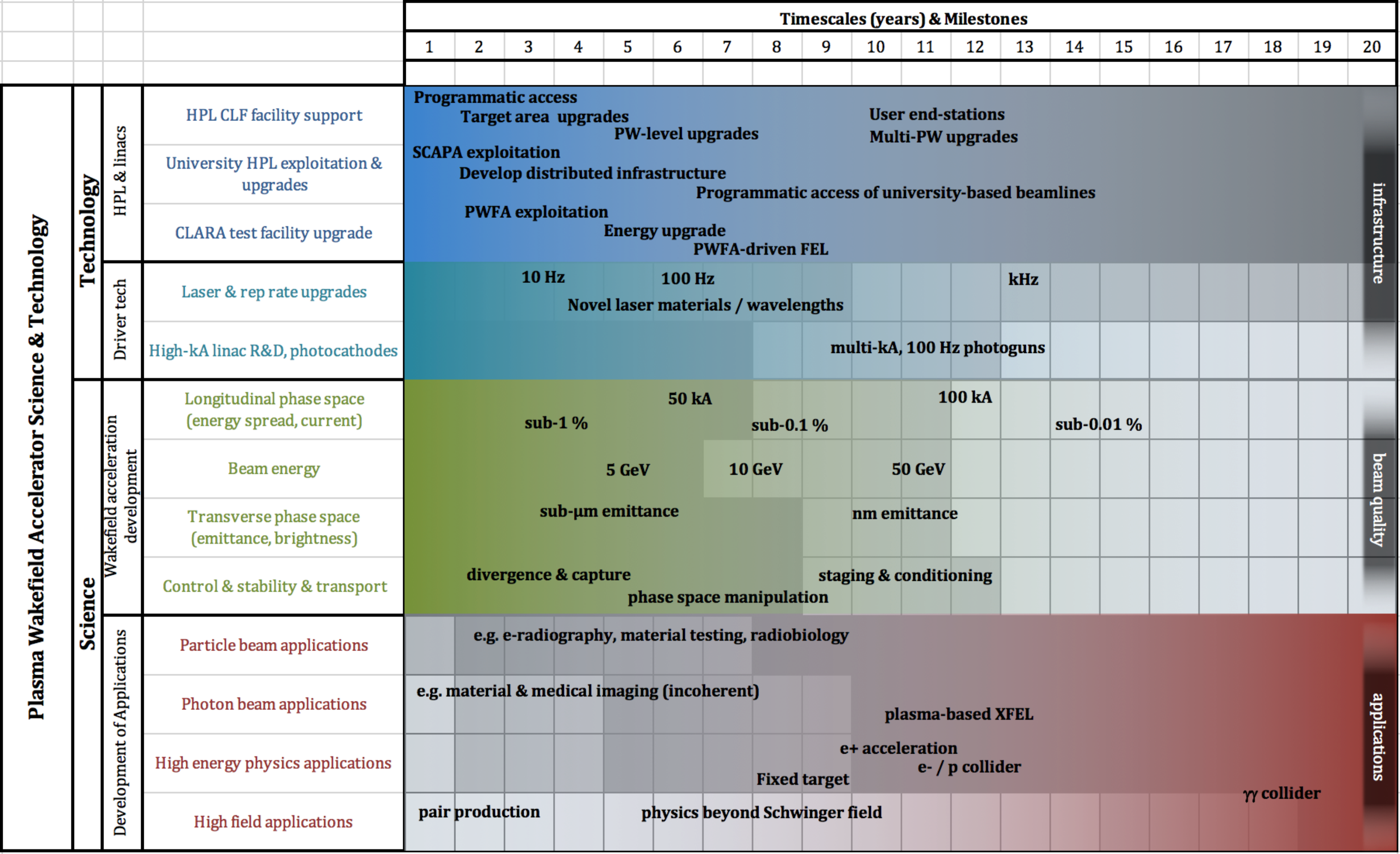}
      \caption{Timeline for scientific and technological research and development on plasma wakefield accelerators.}
      \label{Fig:Timeline}
  \end{figure} 
\end{landscape}

In Fig.\ \ref{Fig:Timeline} we identify major technological and scientific milestones, divided into high-power laser and linac facilities; driver technologies; wakefield accelerator development; and applications. 

It should be noted that the particle beam parameters given in Fig.\ \ref{Fig:Timeline} will not necessarily be available simultaneously or from the same technology. As research progresses the most appropriate approach for delivering each of the key bunch parameters will be established which in turn will determine the most suitable applications of LWFAs and PWFAs.

Achieving the milestones identified in  Fig.\ \ref{Fig:Timeline} will maintain the UK's world-leading position in plasma accelerators. However, doing so will require investment in university groups and national facilites, as described in the following section.

% Required resources
\section{Required resources}
It has been recognised for many years now that ``there is some evidence that the UK laser plasma wakefield accelerator community is losing leadership due to relatively modest investment in this area compared with international competitors in the US and Europe''\cite{STFCAcceleratorReviewReport2014}. This, coupled with the fact that other nations have increased their investments in this area, threatens to result in the UK losing leading its position.

In order to allow the UK to maintain world leadership in this important field, and to meet the scientific challenges identified above, it will be necessary to provide increased support for upgraded infrastructure and facilities, coupled with a strong commitment to training the next generation of scientists in this field. These investments are expected to have high return through industrial exploitation and by driving advances in other scientific disciplines. 

It has previously been suggested that structures for coordinating work in this area should be established.\cite{STFCAcceleratorReviewReport2014} The community has responded e.g. by forming the PWASC, and by developing the current roadmap. These structures could form the basis for efficient planning and management of enhanced investment in this research area.

\subsection{Facilities}\label{Sec:Facilities}
The availability of state-of-the-art laser systems at the CLF has played a vital role in maintaining the UK's world-leading standing in LWFA research.  However, in recent years, a lack of timely upgrades to the facilities has severely limited the rate of development. It is clear that other countries (notably Germany, France, China, US, and Korea) invest far more into plasma wakefield accelerators than the UK: for example, in Germany there are now over 10 laser systems at the \unit[100]{TW} to \unit[1]{PW} level, with funding which includes beam-time costs and programmatic R$\&$D. 

Currently, the preeminent centre for UK LWFA research is the Astra-Gemini laser at the CLF.  The Astra-Gemini laser is one of the most mature tools for LWFA research in the world. However, its world-leading position is severely threatened since: (i) it is a multi-purpose laser system, with only a single target area, which limits its availability for LWFA research to typically only six weeks per year; and (ii) it has not been upgraded for a decade.

The limited available beam time means that the Astra-Gemini laser is significantly oversubscribed. The intense competition for beam time means that, understandably, experiments with potential for immediate high impact are favoured by the Facility Access Panel, and that programmatic work aimed at improving the stability and quality of plasma accelerators is less likely to be given beam time.

In order to achieve significant improvements in the quality and stability of the bunches generated by laser--plasma accelerators it is imperative that the UK community has available dedicated facilities that support plasma accelerator research. We note that these improvements to bunch quality are vital for developing the applications of plasma accelerators, and in due course this investment will lead to high impact results. We note also that many applications of LWFAs, such as betatron emission for high-resolution imaging, require operation at high pulse repetition rates. Hence to move from proof-of-principle tests to real-world applications it will be necessary to increase the laser repetition rate above \unit[10]{Hz} using the diode-pumped solid-state laser (DPSSL) technology.

\pagebreak
Research at CLF is strongly reinforced by university-scale laser centres in Strathclyde (SCAPA), Queen's Belfast (Taranis-X), Oxford, and at ICL. These systems are used for testing novel concepts, preparation for periods of access at national and international facilities, and, importantly, training the next generation of researchers. Investment in university-based centres will therefore be vital for maintaining the health of the research field and to enable sustained progress towards the roadmap goals.

There is currently no facility in the UK for particle-driven plasma accelerator research. So  while the UK groups having leadership roles in large international campaigns and programmes (e.g.\ AWAKE, SLAC FACET), most of the experimental work is naturally undertaken overseas. However, with the appropriate support and development of the CLARA facility in Daresbury Laboratory this situation could  be changed, and in addition to provision of PWFA R\&D opportunities international groups and intellectual output could be attracted .

For the particle-driven plasma accelerator research in the UK, it is important to continue to develop CLARA and establish it as a full-fledged facility, capable of driving \unit[250]{MeV} (or beyond) beams with high currents. The UK should continue to have strong participation in the AWAKE campaign at CERN and at international facilities such as at SLAC FACET-II, DESY FLASHForward and Helmholtz VI, INFN and CERN CLEAR, along with exploiting hybrid schemes with dedicated beam lines at CLARA and SCAPA. 

\begin{recommendation}
\label{Rec:IntlCollaborators}
Mechanisms should be sought to allow UK groups to play leadership roles in international high-visibility collaborations such as SLAC FACET-II, Helmholtz ATHENA, Laserlab Europe, ELI, ARIES etc., and to exploit these.
\end{recommendation}

In the medium term, it is clear that the advancements in this field will require next-generation facilities.  To maintain leadership in this area, we envisage that the UK would require a 100-PW-class laser, along with a \unit[100]{Hz}, PW-class laser facility driving applications in 10--15 years. It is foreseeable that R\&D advancement in this area would enable a plasma-accelerator-based FEL on these timescales. The community strongly believes that a EuPRAXIA-like facility would provide an appropriate staging to de-risk the process.

\subsection{Computation}
Modelling and simulations with high performance computing (HPC) facilities is a large and crucial part of all plasma accelerator R\&D, particularly for particle-in-cell (PIC) simulations. CLF provides a limited access to HPC facilities (SCARF) for users, Daresbury Laboratory offers Hartree, EPSRC manages ARCHER and many universities have their own HPC facilities (with access fees, or with limited free access). However, these opportunities are not universal, and a programmatic, low-threshold access path for Novel Acceleration simulation work should be established in the UK.

A wide range of codes are used by the community, ranging from commercial codes (e.g.\  VORPAL/VSim,  developed by Tech-X) to open-source or self-developed codes. The UK community is particularly fortunate to be able to draw on the use of the EPOCH PIC code for laser/beam/plasma simulations. EPOCH was developed via an EPSRC grant, and is free for academic users. It is currently supported by two EPSRC grants: until September 2019 by grant EP/P02212X/1; and until June 2022 via the Plasma-HEC Consortium (EP/R029148/1), which provides 40\% FTE support for EPOCH. The EPOCH developers also run training workshops (funded by the CLF), which provide training in EPOCH as well in wider HPC skills. Although EPOCH is a mature code, continued development and advances in capability and performance are needed to support the need for more sophisticated and/or more computationally intensive simulations. To date EPOCH support for plasma accelerator work has been undertaken as part of wider code development; direct funding of an EPOCH developer (either partially or fully) to support Novel Accelerator work, would enable the implementation of new capabilities  and faster resolution of bugs or technical problems, and would ensure that this UK code remains internationally competitive.

As noted in \S \ref{Sec:Transport_staging_feedback}, as plasma accelerators mature it becomes increasingly important to develop beam transport systems capable of handling plasma-accelerated beams. The links between groups working on start-to-end simulations of RF-driven systems and those working on plasma wakefield simulations and modelling capabilities should therefore be strengthened; provision of programmatic funding of high-performance computing could foster this.

\subsection{Training, education \& skills}
The university groups, the accelerator institutes (the CI and JAI), and the CLF offer extensive, high-quality training programmes in all aspects of accelerator science which play a vital role in developing the next generation of scientists and engineers. These courses attract a significant number of high quality graduate students from around the world who are an important factor in enabling the UK to maintain an internationally leading position in plasma accelerator research and development.

The UK's training programmes in conventional and novel accelerator science are world leading. Since 2013 more than 100 PhD students have graduated from the CI and JAI. This impressive number is enabled through a total of typically 7--8 STFC quota studentships per annum, plus a similar number of studentships funded by other research councils, scholarships, and research projects. The graduate courses offered by the accelerator institutes cover all aspects of conventional and novel accelerator science, such as beam dynamics, RF cavities, lattice design, magnetic insertion devices, and computational methods --- as well as courses on laser physics, plasma physics, and applications of accelerators. In addition to lectures and classes, students receive training through seminars and design projects.

It should be emphasised that only about 25\% of the accelerator institutes' graduate students work on plasma wakefield accelerators, i.e.\ typically 3--4 graduate students p.a. This number is too small to meet the current level of research activity, and the number of graduate studentships in this area will need to be increased significantly to help realise the opportunity for rapid growth in this research field. 

In many cases a key part of a graduate students work is undertaken at national facilities such as CLF, the Diamond Light Source, VELA, and CLARA. Since the beam time available at these facilities is very limited, university scale facilities play a vital role in training students, as well as allowing them to undertake small-scale studies and to prepare for beam time at a national facility. The role played by in-house facilities could be enhanced by providing a means to allow PhD students to undertake training and short periods of work at UK facilities outside their own institution:

\begin{recommendation}
\label{Rec:NationalScheme}
A national scheme should be developed to enable mobility and knowledge transfer within UK institutions, to increase beam access, and to sustain collaborative efforts; this would provide a means to test new concepts, train students, and prepare for beam time at national and international facilities.
\end{recommendation}

These ambitious developments require increased support for young scientists working in these research areas. However, a current problem is that work on novel accelerators falls between the major areas of expertise on research council fellowship applications panels.  To address this issue we propose the creation of ``Novel accelerator fellowships'' aimed at retaining our most able young scientists in this area of national importance. It is also vital that the number of PhD students working in this area is increased substantially; one mechanism for achieving this could be increasing support for training PhD students by the UK accelerator institutes.

\begin{recommendation}
\label{Rec:Fellowships}
A new "Novel Accelerator Fellowship" scheme should be developed and the support available for training PhD students in novel accelerators should be increased.
\end{recommendation}

\section{Summary}
Laser- and beam-driven plasma accelerators have made rapid advances in recent years, not least as the result of world-leading research by researchers based in the UK. Plasma accelerators can today generate electron beams with GeV-scale energies. They have been used to generate femtosecond-duration pulses of radiation from the visible to hard X-ray wavelengths, which points the way to a new generation of compact, synchronised sources of energetic particles and visible-to-X-ray photons; and they lie at the heart of recent experiments to generate neutral electron--positron plasmas, which offers the enticing prospect of recreating extreme astrophysical environments in the laboratory. In the longer term, plasma accelerators could provide a way to reduce the size and cost of
high-energy particle colliders used at the forefront of physics. It is clear that plasma accelerators have the potential to be a disruptive technology with many potential impacts in science, technology, and medicine.

The UK has high international standing in this field, based on decades of impressive achievements within universities, the accelerator institutes, and at national facilities. The UK continues to be at the forefront of new developments in plasma accelerators, but this position is threatened by a lack of investment in national and university-scale research facilities, as well as by the increase in research investment by other countries.

In this roadmap we have summarised the state-of-the-art, provided a national and international perspective of the field, and outlined the research and development needed further to advance the fields and to develop applications. We have also stated where continued or additional investment will be needed to develop plasma accelerators and their applications.

The plasma accelerator community looks forward to working with the Research Councils, the Government, national facilities, and industry to build on the achievements of UK scientists and realise the enormous promise of this exciting technology.

%Begin Appendices
\newpage
\appendix
\titleformat{\section}{\normalfont\Large\bfseries}{Appendix \thesection}{1em}{}
\titlespacing*{\section}{0em}{2\baselineskip}{0\baselineskip}

\section[Consultation]{Community consultation and roadmap input}\label{Sec:Community_consultation}

The development of this roadmap was initiated by the UK Plasma Wakefield Accelerator Steering Committee (PWASC). This committee was established to represent UK groups working on plasma accelerators, and to help coordinate their activities, and its members are drawn from UK research groups, the Central Laser Facility, and the two Accelerator Science Institutes. 

The PWASC organized a Community Meeting to discuss, and receive feedback on,  the first full draft of the roadmap. This meeting was held on 26th January 2018 and was attended by 30 representatives from universities, national facilities, and industry.

The input received was used to compile a second full draft which was circulated to the community in December 2018 for further comments.

%\newpage
\section{List of abbreviations used}

\vspace{10 mm}

\begin{tabularx}{\linewidth}{lX}
\textbf{Abbreviation} & \textbf{Meaning}\\
\hline
ALEGRO & Advanced LinEar collider study GROup\\
ANA & (ICFA panel on) Advanced and Novel Accelerators\\
AWAKE & Advanced WAKEfield experiment\\
BNL & Brookhaven National Laboratory\\
CALTA & The Centre for Advanced Laser Technology and Applications, Rutherford Appleton Laboratory \\
CI & Cockroft Institute\\
CLARA & the Compact Linear Accelerator for Research and Applications, Daresbury Laboratory\\
CLEAR & CERN Linear Electron Accelerator for Research\\
CLF & Central Laser Facility, Rutherford Appleton Laboratory\\
CSR & Coherent synchrotron radiation \\
DESY & Deutsches Elektronen-Synchrotron\\
DPSSL & Diode-Pumped Solid-State Laser\\
FACET & Facility for Advanced Accelerator Experimental Tests, based at SLAC\\
FEL & Free-electron laser\\
HEP & High Energy Physics\\
HPC & High Performance Computing\\
HPL & High Power Laser\\
ICFA & International Committee for Future Accelerators\\
ICL & Imperial College London\\
ICS & Inverse Compton Scattering\\
INFN & Istituto Nazionale di Fisica Nucleare, Italy\\
JAI & John Adam's Institute for Accelerator Science\\
LWFA & Laser Wakefield Accelerator\\
NEXAFS & Near Edge X-ray Absorption Fine Structure\\
PIC & Particle-In-Cell\\
PWFA & Plasma Wakefield Accelerator\\
QED & Quantum Electrodynamics\\
RBE & Rrelative biological effectiveness\\
RF & Radio-Frequency\\
RT & Radiotherapy \\
SCAPA & Scottish Centre for the Application of Plasma-based Accelerators\\
TDS & Transverse Deflection Structures\\
TRL & Technology Readiness Level\\
VELA & the Versatile Linear Accelerator\\
XANES & X-ray absorption near edge structure\\
\hline
\end{tabularx}
\label{Tab:Abbreviations}

% Bibliography
\newpage
\printbibliography

\end{document}